\documentclass{sig-alternate-sigmod07}

\usepackage{alltt}
\usepackage{latexsym}

\usepackage{complexity}

\usepackage{graphics}
\usepackage{epsf}
\usepackage{epsfig}

\usepackage{amsgen}
\usepackage{amstext}
\usepackage{amsmath}
\usepackage{amsfonts}
\usepackage[mathscr]{euscript}
\usepackage{bbm}
\usepackage{algorithmic}
\usepackage[section]{algorithm}

\usepackage{moreverb}
\usepackage{latexsym}
\usepackage{amsgen}
\usepackage{amsmath}
\usepackage{amstext}
\usepackage{amscd}
\usepackage{amssymb}
\usepackage{enumerate}

\def \Pr{\mathbf{P}}

\newtheorem{theorem}{Theorem}[section]          
\newtheorem{lemma}[theorem]{Lemma}              
\newtheorem{corollary}[theorem]{Corollary}      
          
\newtheorem{definition}[theorem]{Definition}    
\newtheorem{proposition}[theorem]{Proposition}  
 \newenvironment{example}                        
   {\refstepcounter{theorem}\trivlist\item       
   [\hskip\labelsep\bf Example \thetheorem]}
   {\endtrivlist}                                

\newcommand{\set}[1]{\{ #1 \}}

\renewcommand{\vec}[1]{\mbox{\boldmath$#1$}}

\newcommand{\eat}[1]{}
\newcommand{\ND}{\mbox{\it ND}}
\newcommand{\MD}{\mbox{\it MD}}
\renewcommand{\UP}{\texttt{up}}
\newcommand{\ceil}[1]{{\lceil #1 \rceil}}
\renewcommand{\IP}{\texttt{ip}}
\newcommand{\LP}{\texttt{lp}}
\newcommand{\factors}{\mbox{\it Factors}}

\newcommand{\jp}{{\mbox{\it jp}}}
\newcommand{\ip}{{\mbox{\it ip}}}

  {\refstepcounter{theorem}\trivlist\item       
  [\hskip\labelsep\bf\sf Policy Query \thetheorem]}
  {\endtrivlist}                                

  {\displaymath\def\\{{\ifnum0=`}\fi\def\\####1
  {\def\tmpa{####1}\def\tmpb{[}\ifx\tmpa\tmpb   
  \def\\########1]{\ifnum0=`{\fi}\hidewidth\cr  
  \noalign{\ifdim########1<1ex\penalty100\fi    
  \vskip########1\vskip\jot}}\relax             
  \else\def\\{\ifnum0=`{\fi}\hidewidth\cr       
  \noalign{\penalty100\vskip\jot}####1}\fi\\}\\}
  \tabskip=1em\halign\bgroup\tabskip.3em&\vbox  
  {\hrule height0pt width.7em\hbox{$##$}}\cr}
  {\hidewidth\cr\egroup\enddisplaymath}         
\newcommand{\setof}[2]{\{{#1}\mid{#2}\}}        
\mathcode`\?="426E                              
\mathcode`\!="005E                              
\mathcode`\*="0203                              
\newcommand{\makeop}[2]                         
  {\ifx#2.\def\next##1{}\else\escapechar=-1     
  \def\next##1{\escapechar=92\def#2{#1}}        
  \expandafter\next\expandafter{\string#2}      
  \let\next\makeop\fi\next{#1}}                 
\makeop{{\mskip-5mu}\mathrel{\sf{\lowercase{#1}}}}
   \Input \Output \Select \Where \Range \Create \Link \Collect  %
   \If  \Foreach \Let .                         %
\makeop{\mathrel{\sf{\lowercase{#1}}}}          
  \From \Case \Of \And \Then \Else \Giving \appr
  \DB \IN \Not \matches \Do \In \flatten .              %
\makeop{{\sl #1\/}}                             
  \String \Int \Real \Symbol                    %
  \Tree \Label \Oid \Bool \Dom .                %
\makeop{\mathop{\sl\lowercase{#1}\/}\nolimits}  
  \isstring \isempty \deepflatten               %
  \gext \ext \vext \nodes \edges \root          %
  \false \true                                  %
  \Set \nest \unnest \joinnest                  %
  \outernest .                                  %
\makeop{{\cal #1}}                              
  \C \B \D \F \G \H \U \V \W \X \Y \T \Z \D \I \C \A \O \L.    %

\newcommand{\itb}[1]{\mbox{\it #1}}

\newcommand{\summation}{{\mathbf \bigoplus}}

\def\AddSpace#1{\ifcat#1a\ \fi#1} 

\newcommand{\var}{\itb{var}}

\def \up(#1){[#1)}
\def \down(#1){(#1]}

\def \sec(#1,#2){#1 \: | \: #2}
\def \secd(#1,#2,#3){#1 \: |_{#2} \: #3}
\def \secdp(#1,#2,#3,#4){#1 \: |_{#2,#3} \: #4}
\def \usec(#1,#2,#3){\set{\series(#1,#2)} \:|\: #3}

\def \var(#1){{\bf #1}}

\def \Pr{\mathbf{P}}

\def \series(#1,#2){#1_1, \dots \; #1_{#2}}
\def \serieszero(#1,#2){#1_0, #1_1, \dots \; #1_{#2}}

\def \para(#1){{\vspace{1ex}\noindent\small\bf #1\hspace{1ex}}}
\def \myem(#1){{\vspace{1ex}\noindent\small\em #1\hspace{1ex}}}

\renewcommand{\co}{\text{\em coeff}}

\newcommand{\struct}[1]{\mbox{$\bf #1$}}

\begin{document}

\title{The Dichotomy of Conjunctive Queries on Probabilistic Structures}

\author{Nilesh Dalvi and Dan Suciu\\University of Washington, Seattle.}

\maketitle

\begin{abstract}
We show that for every conjunctive query, the complexity of evaluating
it on a probabilistic database is either \PTIME\ or \#\P-complete, and
we give an algorithm for deciding whether a given conjunctive query is
\PTIME\ or \#\P-complete. The dichotomy property is a fundamental
result on query evaluation on probabilistic databases and it gives a
complete classification of the complexity of conjunctive queries.

\end{abstract}

\section{Problem Statement}

\label{sec:overview}

Fix a relational vocabulary $R_1, \ldots, R_k$, denoted ${\cal R}$.  A
{\em tuple-indepen\-dent probabilistic structure} is a pair $({\bf A},
p)$ where ${\bf A}$ = ($A$, $R_1^A$, $\ldots$, $R_k^A$) is first order
structure and $p$ is a function that associates to each tuple $t$ in
${\bf A}$ a rational number $p(t) \in [0,1]$.  A probabilistic
structure $({\bf A}, p)$ induces a probability distribution on the
set of substructures ${\bf B}$ of ${\bf A}$ by:
\begin{eqnarray}
p({\bf B}) & = & \prod_{i=1}^k (\prod_{t \in R^B_i} p(t) \times
\prod_{t \in R^A_i - R^B_i} (1-p(t)))
  \label{eq:pwd}
\end{eqnarray}
\noindent where ${\bf B} \subseteq {\bf A}$, more precisely ${\bf B} =
(A, R_1^B, \ldots, B_k^B)$ is s.t. $R_i^B \subseteq R_i^A$ for $i = 1,
k$.

A {\em conjunctive query}, $q$, is a sentence of the form $\exists
\bar x.  (\varphi_1 \wedge \ldots \wedge \varphi_m)$, where each
$\varphi_i$ is a positive atomic predicate $R(t)$, called a {\em
  sub-goal}, and the tuple $t$ consists of variables and/or constants.
As usual, we drop the existential quantifiers and the $\wedge$,
writing $q = \varphi_1, \varphi_2, \ldots, \varphi_m$.  A {\em conjunctive property} is a
property on structures defined by a conjunctive query $q$, and its
probability on a probabilistic structure $({\bf A}, p)$ is defined as:
\begin{eqnarray}
p(q) & = & \sum_{{\bf B} \subseteq {\bf A}: {\bf B} \models q} p({\bf B})
  \label{eq:prq}
\end{eqnarray}
In this paper we study the data complexity of Boolean conjunctive
properties on tuple independent probabilistic structures.  (When clear
from the context we blur the distinction between queries and
properties).
\eat{
More precisely, for a fixed vocabulary and a Boolean
conjunctive query $q$ we study the following two problems:
\begin{description}
\item[{\sc Evaluation}] For a given probabilistic structure $({\bf A}, p)$,
  compute the probability $p(q)$.
\item[{\sc Counting}] For a given structure ${\bf A}$, define $p(t) = 1/2$
  for every tuple $t$ and compute $p(q)$.
\end{description}
The {\sc Counting} problem counts, for a give structure ${\bf A}$, the
number of substructures of ${\bf A}$ that satisfy the given query.  In
both cases the complexity is in the size of ${\bf A}$ and (for the
evaluation problem) in the size of the representations of the rational
numbers $p(t)$.  Both problems are trivially contained in \#\P.  We
show conditions under which the evaluation problem is in \PTIME, and
conditions under which the counting problem is \#\P-hard.  In the
first case we say that ``the query is in \PTIME'', while in the second
we say that ``it is \#\P-hard''. Our main result in this paper is the
following:

\begin{theorem} (Dichotomy Theorem) Given any conjunctive query $q$,
the complexity of {\sc Evaluation} and {\sc Counting} are either
\PTIME\ or \#\P-complete. The problem of deciding if a give query is
\PTIME\ is $\Sigma_2^P$-complete in the size of the query.
\end{theorem}
}

More precisely, for a fixed vocabulary and a Boolean conjunctive query
$q$ we study the following problem:
\begin{description}
\item[{\sc Evaluation}] For a given probabilistic structure $({\bf A}, p)$,
  compute the probability $p(q)$.
\end{description}
The complexity is in the size of ${\bf A}$ and in the size of the
representations of the rational numbers $p(t)$.  This problem is
trivially contained in \#\P , and we show conditions under which it
is in \PTIME, and conditions where it is \#\P-hard. The class \#\P
~\cite{valiant:sharpP} is the counting analogue of the class $\NP$.

\begin{theorem} (Dichotomy Theorem) Given any conjunctive query $q$,
  the complexity of {\sc Evaluation} is either \PTIME\ or
  \#\P-complete.
\end{theorem}

{\bf Background and motivation} Dichotomy theorems are fundamental to
our understanding of the structure of conjunctive queries. A widely
studied problem, which can be viewed as the dual of our problem, is
the {\em constraint satisfaction problem} ({\sc CSP}) and is as follows:
given a fixed relational structure, what is the complexity of
evaluating conjunctive queries over the structure?
Shaefer~\cite{schaefer78dichotomy} has shown that over binary domains,
{\sc CSP} has a dichotomy into \PTIME\ and \NP-complete. Feder and
Vardi~\cite{vardi93dichotomy} have conjectured that a similar dichotomy
holds for arbitrary (non-binary) domains. Creignou and
Hermann~\cite{hermann96counting} showed that the counting version of
the {\sc CSP} problem has a dichotomy into \PTIME\ and \#P-complete. The
problem we study in this paper seems different in nature, yet still
interesting.

In addition to the pure theoretical interest we also have a practical
motivation.  Probabilistic databases are increasingly used to manage a
wide range of imprecise data~\cite{trio,mystiq}.  But general
purpose probabilistic database are difficult to build, because query
evaluation is difficult: it is both theoretically hard
(\#\P-hard~\cite{gradel:reliability,dalvi04prob}) and plain difficult
to understand.  All systems reported in the literature have
circumvented the full query evaluation problem by either severely
restricting the queries~\cite{barbara:probdb}, or by using a
non-scalable (exponential) evaluation algorithm~\cite{fuhr:tis:97}, or
by using a weaker semantics based on intervals~\cite{laks:probview}.
In our own system, MystiQ~\cite{mystiq}, we support arbitrary
conjunctive queries as follows.  For queries without self-joins, we
test if they have a \PTIME\ plan using the techniques
in~\cite{dalvi06debul}; if not, then we run a Monte Carlo
simulation algorithm.  The query execution times between the two cases
differ by one or two orders of magnitude (seconds v.s. minutes).  The
desire to improve MystiQ's query performance on arbitrary queries
(i.e. with self-joins) has partially motivated this work.

\subsection{Overview of Results}

\label{subsec:overview}

We summarize here our main results on the query evaluation
problem.  Some of this discussion is informal and is intended to
introduce the major concepts needed to understand the evaluation of
conjunctive queries on probabilistic structures.

{\bf Hierarchical queries:} For a conjunctive query $q$, let $Vars(q)$
denote its set of variables, and, for $x \in Vars(q)$, let $sg(x)$ be
the set of sub-goals that contain $x$.
\begin{definition} A conjunctive query is  {\em hierarchical} if for
  any two variables $x, y$, either $sg(x) \cap sg(y) = \emptyset$, or
  $sg(x) \subseteq sg(y)$, or $sg(y) \subseteq sg(x)$.  We write $x
  \sqsubseteq y$ whenever $sg(x) \subseteq sg(y)$ and write $x \equiv
  y$ when $sq(x) = sg(y)$.  A conjunctive property is hierarchical if
  it is defined by some hierarchical conjunctive query.
\end{definition}
It is easy to check that a conjunctive property is hierarchical if the
minimal conjunctive query defining it is hierarchical. As an example,
the query $q_{\mbox{\scriptsize hier}} = R(x), S(x,y)$ is hierarchical
because $sg(x) = \set{R, S}$, $sg(y) = \set{S}$. On the other hand,
the query $q_{\mbox{\scriptsize non-h}} = R(x), S(x,y), T(y)$ is not
hierarchical because $sg(x) = \set{R, S}$ and $sg(y) = \set{S,T}$.
\eat{
\PTIME\footnote{$p(q) = 1 - \prod_{a \in
    A}(1-p(R(a))(1-\prod_{b \in A}(1- p(S(a,b)))))$}
}

In prior work~\cite{dalvi04prob} we have studied the evaluation
problem under the following restriction: every sub-goal of $q$ refers
to a different relation name.  We say that $q$ has no self-joins.  The
main result in~\cite{dalvi04prob}, restated in the terminology used here, is:
\begin{theorem} \label{th:old} \cite{dalvi04prob} Assume $q$ has no
  self joins.  Then: (1) If $q$ is hierarchical, then it is in \PTIME.
  (2) If $q$ is not hierarchical then it is \#\P-hard.
\end{theorem}
Moreover, the \PTIME\ algorithm for a hierarchical query is the
following simple recurrence on query's structure.  Call a variable $x$
{\em maximal} if for all $y$, $y \sqsupseteq x$ implies $x \sqsupseteq
y$.  Pick a maximal variable from each connected component of the
query to obtain the set $x_1, \ldots, x_m$.  Let $f_0, f_1(x_1), \ldots, f_m(x_m)$ be the connected
components of $q$: $f_0$ contains all constant sub-goals, and
$f_i(x_i)$ consists of all sub-goals containing $x_i$ for $i =1,m$.
Then:
\begin{eqnarray}
  p(q) & = & p(f_0) \cdot \prod_{i=1,m}(1 - \prod_{a \in  A}(1-p(f_i[a/x_i]))) \label{eq:safeplan}
\end{eqnarray}
This formula is a recurrence on the query's structure (since each
$f_i[a/x_i]$ is simpler than $q$) and it is correct because
$f_i[a/x_i]$ is independent from $f_j[a'/x_j]$ whenever $i\not=j$ or
$a\not=a'$. As an example, for query $q_{\mbox{\scriptsize hier}} =
R(x), S(x,y)$, $p(q) = 1 - \prod_{a \in A}(1-p(R(a))(1-\prod_{b \in
A}(1- p(S(a,b)))))$.

In this paper we study arbitrary conjunctive queries (i.e. allowing
self-joins), which turn out to be significantly more complex.  The
starting point is the following extension of Theorem~\ref{th:old} (2)
(the proof is in the appendix):
\begin{theorem} \label{th:non-hierarchical} If $q$  is not
  hierarchical then it is \#\P-hard.
\end{theorem}
Thus, from now on we consider only hierarchical conjunctive queries in
this paper, unless otherwise stated.

{\bf Inversions: } As a first contact with the issues raised by
self-joins, let us consider the following query:
\begin{eqnarray*}
q & = & R(x),S(x,y), S(x',y'), T(x')
\end{eqnarray*}
\noindent We write it as $q = f_1(x) f_2(x')$, where $f_1(x) =
R(x),S(x,y)$ and $f_2(x') = S(x',y'),T(x')$.  The query is
hierarchical, but it has a self-join because the symbol $S$ occurs
twice: as a consequence $f_1[a/x]$ is no longer independent from
$f_2[a/x']$ (they share common tuples of the form $S(a,b)$), which
prevents us from applying Equation (\ref{eq:safeplan}) directly.  Our
approach here is to define a new query by equating $x=x'$, $f_3(x) =
f_1(x)f_2(x) = R(x),S(x,y),S(x,y'),T(x)$ which is equivalent to
$R(x),S(x,y),T(x)$.  We show that the probability $p(q)$ can be
expressed using recurrences over the probabilities of queries of the
form $f_1[a/x_1]$, $f_2[a'/x_2]$, $f_3[a''/x_3]$, as a sum of a few
formulas\footnote{This particular example admits an alternative,
  perhaps simpler PTIME solution, based on a dynamic programming
  algorithm on the domain $A$.  For other, very simple queries, we are
  not aware of any algorithm that is simpler than ours (formula
  (\ref{eq:prob-recur}), Sec.~\ref{sec:noinversions-ptime}), for
  example $R(x,y,y,x),R(x,y,x,z)$, or
  $R(y,x,y,x,y),R(y,x,y,z,x),R(x,x,y,z,u)$ (both are in PTIME because
  they have no inversions). To appreciate the difficulties even with
  such simple queries note that, by contrast,
  $R(y,x,y,x,y),R(y,y,y,z,x),R(x,x,y,z,u)$ is \#\P-hard. For
  additional challenging PTIME queries, see
  Fig.~\ref{fig:inversion-free}.} in the same style as
(\ref{eq:safeplan}) (see Example~\ref{ex:no-inversion-ptime}).  The
correctness is based on the fact that $f_i[a/x_i]$ and $f_j[a'/x_j]$
are independent if $i\not=j$ or $a \not=a'$.

However, this approach fails when the query has an ``inversion''.
Consider:
\begin{eqnarray*}
H_0 & = & R(x),S(x,y), S(x',y'),T(y')
\end{eqnarray*}
\noindent This query is hierarchical, but the above approach no longer
works.  The reason is that the two sub-goals $S(x,y)$ and $S(x',y')$
unify, while $x \sqsupset y$ and $x' \sqsubset y'$: we call this an
{\em inversion} (formal definition is in
Sec.~\ref{sec:inversion-definition}).  If we write $H_0$ as
$f_1(x)f_2(y')$ and attempt to apply a recurrence formula, the queries
$f_1[a/x]$ and $f_2[a'/y']$ are no longer independent even if $a \not=
a'$, because they share the common tuple $S(a,a')$.

Inversions can occur as a result of a chain of unifications:
  \begin{tabbing}
    $H_k$ =\\
    $      R(x),$\=$S_0(x,y),$ \\
     \>$S_0(u_1,v_1),$\=$S_1(u_1,v_1)$ \\
     \>       \>$S_1(u_2,v_2)$,\ldots\= \\
     \>       \>      \>$S_{k-1}(u_k,v_k),$\=$S_k(u_k,v_k)$ \\
     \>       \>      \>     \>$S_k(x',y'),T(y')$
  \end{tabbing}
\noindent Here any two consecutive pairs of variables in the sequence $x
\sqsupset y$, $u_1 \equiv v_1$, $u_2 \equiv v_2$, \ldots, $x'
\sqsubset y'$ unify, and we also call this an inversion.  We prove in
the Appendix:
\begin{theorem} \label{th:hard}
  For every $k \geq 0$, $H_k$ is \#\P-hard.
\end{theorem}
Thus, some hierarchical queries with inversions are \#\P-hard.  We
prove, however, that if $q$ has no inversions, then it is in \PTIME:

\eat{We show this by proving that, after some appropriate rewriting of
  the query, there exists an order relation $\succeq$ on the variables
  that refines the preorder $\sqsupseteq$ s.t. whenever a pair of
  variables $x,y$ unifies with $x',y'$ and $x \succ y$, then $x' \succ
  y'$: we call $\succeq$ a {\em safe} order.  Using $\succ$ we derive
  a recurrence formula generalizing (\ref{eq:safeplan}) that computes
  the query in PTIME:}

\begin{theorem} \label{th:no-inversions}
  If $q$ is hierarchical and has no inversions, then it is in \PTIME.
\end{theorem}

The \PTIME\ algorithm for inversion-free queries is a sum of
recurrence formulas, each similar in spirit to (\ref{eq:safeplan}).
The proof is in Sec.~\ref{sec:noinversions-ptime}.

{\bf Erasers} The precise boundary between \PTIME\ and \#\P-hard
queries is more subtle than simply testing for inversions: some
queries with inversion are \#\P-hard, while others are in \PTIME, as
illustrated below:

\begin{example}
\label{ex:eraserquery}
Consider the hierarchical query $q$
\begin{tabbing}
  $q = $\=$R(r,x),$\=$S(r,x,y),U(a,r),U(r,z),V(r,z)$ \\
        \>         \>$S(r',x',y'),T(r',y'),V(a,r')$ \\
        \>$R(a,b), S(a,b,c), U(a,a)$
\end{tabbing}
\end{example}

\noindent Here $a,b,c$ are constants and the rest are variables.  This
query has an inversion between $x \sqsupset y$ and $x' \sqsubset y'$
(when unifying $S(r,x,y)$ with $S(r',x',y')$).  Because of this
inversion, one may be tempted to try to prove that it is \#\P-hard,
using a reduction from $H_0$.  Our standard construction 
starts by equating $r = r'$ to make $q$ ``like'' $H_0$: call $q'$
the resulting query (i.e. $q' = q[r/r']$).  If one works out the
details of the reduction, one gets stuck by the existence of the
following homomorphism from $h : q \rightarrow q'$ that ``avoids the
inversion'': it maps the variables $r, x, y, z, r', x', y'$ to $a, b,
c, r, r, x', y'$ respectively, in particular sending $U(r,z), V(r,z)$
to $U(a,r)$, $V(a,r)$.  Thus, $h$ takes advantage of the two sub-goals
$U(a,r)$, $V(a,r)$ in $q'$ which did not exists in $q$, and its image
does not contain the sub-goal $S(r,x,y)$, which is part of the
inversion.  We call such a homomorphism an {\em eraser} for this
inversion: the formal definition is in Sec.~\ref{sec:expansion}.
Because of this eraser, we cannot use the inversion to prove that the
query is \#\P-hard.  So far this discussion suggests that erasers are
just a technical annoyance that prevent us from proving hardness of
some queries with inversions.  But, quite remarkably, erasers can also
be used in the opposite direction, to derive a \PTIME\ algorithm: they
are used to cancel out (hence ``erase'') the terms in a certain
expansion of $p(q)$ that correspond to inversions and that do not have
polynomial size closed forms.  Thus, our final result (proven in
Sections~\ref{sec:ptime} and \ref{sec:inversion-hard}) is:

\begin{theorem}[Dichotomy] \label{th:dichotomy} Let $q$ be hierarchical. \\
  (1) If $q$ has an inversion without erasers then $q$ is \#\P-hard.
  \newline (2) If all inversions of $q$ have erasers then $q$ is in
  \PTIME.
\end{theorem}

As a non-trivial application of (1) we show (Fig.~\ref{fig:inversion}
in Appendix~\ref{app:examples} and in Example~\ref{ex:hardness1})
that each of the following two queries are \#\P-hard, since each has
an inversion between two isomorphic copies of itself:
\begin{eqnarray*}
  q_{\mbox{\scriptsize 2path}} & = & R(x,y), R(y,z) \\
  q_{\mbox{\scriptsize marked-ring}} & = & R(x),S(x,y),S(y,x)
\end{eqnarray*}
\noindent In general, the hardness proof  is by reduction from the query
$H_k$, where $k$ is the length of an inversion without an eraser.  The
proof is not straightforward.  It turns out that not every eraser-free
inversion can be used to show hardness.  Instead we show that if there
is an eraser-free inversion then there is one that admits a reduction
from $H_k$. 
\eat{
Finally, we point out that the hardness reductions for \#\P-hard
queries are for the weaker flavor of hardness, where the probabilities
are given as arbitrary rational numbers. It would be desirable to show
that all \#\P-hard queries are also hard when all probabilities are
fixed to $1/2$ (sometimes called the {\em counting problem}): we leave
this for future work.
}

The \PTIME\ algorithm in (2) is also not straightforward at all.  It
is quite different from the recurrence formula in
Theorem~\ref{th:no-inversions}, since we can no longer iterate on the
structure of the query: in Example ~\ref{ex:eraserquery}, the sub-query of $q$ consisting
of the first two lines is \#\P-hard (since without the third line
there is no eraser), hence we cannot compute it separately from the
third line.  Our algorithm here computes $p(q)$ without recurrence,
and thus is quite different from the inversion-free \PTIME\ algorithm,
but uses the latter as a subroutine.

\eat{
Finally, we note that Theorems~\ref{th:non-hierarchical},
\ref{th:no-inversions}, \ref{th:dichotomy} prove a dichotomy for
conjunctive queries on probabilistic structures.
}

\eat{
{\bf The Hard Queries} For any hard conjunctive query there is a
reduction either from $q_{\mbox{\scriptsize non-h}}$ of from some
$H_k$, for $k \geq 0$, and for the latter two types of queries we
prove hardness directly (using a reduction from counting of
partitioned 2DNF~\cite{hermann96counting}, which is quite non-trivial
for $H_k$).  The proofs for both $q_{\mbox{\scriptsize non-h}}$ and
$H_k$ are for the weaker flavor of hardness, where the probabilities
are given as arbitrary rational numbers.  It would be desirable to
show that all \#\P-hard conjunctive queries are also hard when all
probabilities are fixed to $1/2$ (sometimes called the {\em counting
  problem}): we leave this for future work.
}

\section{An Expansion Formula for Conjunctive Queries}

In this section, we introduce the key terminology and prove an
expansion formula for computing the probability of conjunctive
queries that will be used to device \PTIME\ algorithms for query
evaluation.  For the remainder of the paper, all queries are assumed
to be hierarchical, as we know that non-hierarchical queries are
\#\P-hard (Appendix~\ref{app:nonhierarchical}).

\eat{
Both the formulation of the algorithm and the definition of inversions
require that every unifiers between two sub-goals be ``strict'', i.e.
map variables in one sub-goal one to one to the variables in the other
sub-goal.  In general a query may have non-strict unifiers, and to
ensure their strictness we introduce arithmetic predicates $<$ and/or
$\not=$ in the query.
}

\subsection{Coverage}

We call an {\em arithmetic predicate} a predicate of the form $u=v$,
$u \not=v$, or $u < v$ between a variable and a constant in $C$, or
between two variables\footnote{As usual we require every variable to
  be range restricted, i.e. to occur in at least one sub-goal.}. A {\em
  restricted arithmetic predicate} is an arithmetic predicate that is
either between a variable and a constant, or between two variables
$u,v$ that co-occur in some sub-goal (equivalently $u \sqsupseteq v$ or
$u \sqsubseteq v$). From now on, we will allow all conjunctive queries
to have restricted arithmetic predicates.

\begin{definition} 
A {\em coverage} for a query $q$ is a set of conjunctive queries 
$\C = \set{qc_1, \ldots, qc_n}$ such that:
\begin{eqnarray*}
q & \equiv & qc_1 \vee \ldots \vee qc_n
\end{eqnarray*}
Each query in $\C$ is called a {\em cover}.
A {\em factor} of $\C$ is a connected component of some $qc_i \in C$.
We denote the set of all factors in $\C$ by $\F = \set{f_1, \ldots,
f_k}$. 

We alternatively represent a coverage by the pair $(\F, C)$,
where $\F$ is a set of factors and $C$ is a set of subsets of $\F$.
Each element of $C$ determines a cover consisting of the corresponding
set of factors from $\F$. 
\end{definition}

\eat{
We will assume w.l.o.g. that the coverage is non-redundant, i.e.
whenever $qc_i \subseteq qc_j$ we have $i=j$: otherwise, we can remove
$qc_i$ from $\C$ and still have a coverage.

NOTE: We need coverages that are redundant
}

For any query $q$ the set $\C = \set{q}$ is a trivial coverage. We
also define $\C^<(q)$, which we call the {\em canonical coverage},
obtained as follows.  Consider all $m$ pairs $(u,v)$ of co-occurring
variables $u,v$ in $q$, or of a variable $u$ and constant $v$.  For
each such pair choose one of the following predicates: $u<v$ or $u=v$
or $u>v$, and add it to $q$.  This results in $3^m$ queries.  Remove
the unsatisfiable ones, then remove all redundant ones (i.e. remove
$qc_i$ if there exists another $qc_j$ s.t. $qc_i \subset
qc_j$).  The resulting set $\C^<(q) = \set{qc_1, \ldots, qc_n}$ is the
{\em canonical coverage} of $q$.
\\\\
\noindent{\bf Unifiers}
\\
Let $q, q'$ be two queries (not necessarily distinct).  We rename
their variables to ensure that $Vars(q) \cap Vars(q') = \emptyset$,
and write $qq'$ for their conjunction.  Let $g$ and $g'$ be two
sub-goals in $q$ and $q'$ respectively.  The {\em most general
  unifier}, MGU, of $g$ and $g'$ (or the MGU of $q, q'$ when $g, g'$
are clear from the context) is a substitution $\theta$ for $qq'$ s.t.
(a) $\theta(g) = \theta(g')$, (b) for any other substitution $\theta'$
s.t. $\theta'(g)=\theta'(g')$ there exists $\rho$ s.t. $\rho \circ
\theta = \theta'$.

A {\em 1-1 substitution} for queries $q,q'$ is a substitution $\theta$ for
$qq'$ such that:
  (a) for any variable $x$ and constant $a$ $\theta(x) \not=a$, and
  (b) for any two distinct variables $x,y$ in $q$ (or in $q'$), $\theta(x)
  \not= \theta(y)$.
The {\em set representation} of a 1-1 substitution $\theta$ is the set
$\set{(x,y) \mid x \in Vars(q), y \in Vars(q'), \theta(x)=
\theta(y)}$.

\begin{definition}
  An MGU $\theta$ for two queries $q, q'$ is called {\em strict} if
  it is a 1-1 substitution for $qq'$.
\end{definition}

For a trivial illustration, if $q = R(x,x,y,a,z)$ and
$q'=R(u,v,v,w,w)$ and their MGU is $\theta$, then
$\theta(x)=\theta(y)=\theta(u)=\theta(v) = x'$, $\theta(w) = \theta(z)
= a$, and the effect of the unification is $\theta(qq') =
R(x',x',x',a,a)$.  This is not strict: e.g.  $\theta(x) = \theta(y)$
and also $\theta(z) = a$. We want to ensure that all unifications are
strict.

\begin{definition} (Strict coverage)
  Let $\C$ be a coverage and $\F$ be its factors.  We say that $\C$ is
  {\em strict} if any MGU between any two factors $f, f' \in \F$ is
  strict.
\end{definition}

\begin{example}
  Let $q = T(x), R(x,x,y), R(u,v,v)$. The trivial coverage $\C =
  \set{q}$ is not strict, as the $MGU$ of the two $R$ sub-goals of $q$
  equate $x$ with $y$ and $u$ with $v$. Alternatively, consider the
  following three queries:
\begin{eqnarray*}
qc_1 & = & T(x), R(x,x,x)\\
qc_2 & = & T(x), R(x,x,y), R(u,u,u), x \neq y\\
qc_3 & = & T(x), R(x,x,y), R(u,v,v), x \neq y, u \neq v
\end{eqnarray*}
One can show that $q \equiv qc_1 \vee qc_2 \vee qc_3$, hence $\C =
\set{qc_1, qc_2, qc_3}$ is a coverage for $q$. The set of factors
$\F$ consists of the connected components of these queries, which are 
$$
\begin{array}{rclrcl}
f_1 &=& T(x),R(x,x,x) &
f_2 &=& T(x),R(x,x,y),x \neq y\\
f_3 &=& R(u,u,u) &
f_4 &=& R(u,v,v), u \neq v
\end{array}
$$
and $C = \set{\set{f_1},\set{f_2,f_3},\set{f_2,f_4}}$. The coverage
is strict, as a unifier cannot equate $x$ with $y$ or $u$ with $v$ in
any query because of the inequalities. Similarly, the canonical
coverage $\C^<(q)$, which has nine covers containing combinations of $x
< y$, $x = y$, or $x > y$ with $u < v$, $u = v$, $u > v$, is also strict.
\end{example}

\begin{lemma}
The canonical coverage $\C^<(q)$ is always strict.
\end{lemma}

\subsection{Inversions}
\label{sec:inversion-definition}

Fix a strict coverage $\C$ for $q$, with factors $\F$, and define the
following undirected graph $G$.  Its nodes are triples $(f, x, y)$
with $f \in \F$ and $x,y \in Vars(f)$, and its edges are pairs
$((f,x,y),(f',x',y'))$ s.t. there exists two sub-goals $g, g'$ in
$f, f'$ respectively whose MGU $\theta$ satisfies $\theta(x) =
\theta(x')$ and $\theta(y) = \theta(y')$.  We call an edge in $G$ a
{\em unification edge}, and a path a {\em unification path}.  Recall
that for a preorder relation $\sqsupseteq$, the notation $x \sqsupset
y$ means $x \sqsupseteq y$ and $x \not\sqsubseteq y$.

\begin{definition} (Inversion-free Coverage)
  An inversion in $\C$ is a unification path from a node $(f, x, y)$
  with $x \sqsupset y$ to a node $(f', x', y')$ with $x' \sqsubset
  y'$. An {\em inversion-free coverage} is a strict coverage that does
  not have an inversion. We say that $q$ is inversion-free if it has
  at least one inversion-free coverage. Otherwise, we say that $q$ has
  inversion.
\end{definition}

Obviously, to check whether $\C$ has an inversion it suffices to look
for a path in which all intermediate nodes are of the form $(f'',
u,v)$ with $u \equiv v$, i.e. the $\sqsupset$ and $\sqsubset$ are only
at the two ends of the path. The following result says that to check
if a query has an inversion, it is enough to examine the canonical
coverage.

\eat{
Consider $q = R(x,x,y),R(u,v,v)$.  Then the trivial coverage $\C =
\set{q}$ is not strict, while the canonical coverage $\C^<$ (with 9
covers) is strict (since no two sub-goals unify in $\C^<$).  In
general:
}

\begin{proposition}
If there exists one coverage of $q$ that does not contain inversion,
then the canonical cover $\C^<(q)$ does not contain inversion.
\end{proposition}

\begin{example}  We illustrate with two examples:

  (a) Consider $H_k$ in Theorem~\ref{th:hard}.  The trivial coverage
    $\C = \set{H_k}$ is strict, and has factors $\F = \set{f_0, f_1,
      \ldots, f_{k+1}}$ (each line in the definition of $H_k$ is one
    factor).  The following is an inversion: $(f_0, x, y)$, $(f_1,
    u_1, v_1)$, \ldots, $(f_k, u_k, v_k)$, $(f_{k+1}, x', y')$.  This
    is an inversion because $x \sqsupset y$ and $x' \sqsubset y'$.
    The canonical coverage $\C^<$ also has an inversion, e.g. along
    the factors obtained by adding the predicates $x < y$, $u_1 <
    v_1$, \ldots, $u_k < v_k$, $x' < y'$.

    (b) Consider the query $q = R(x), S(x,y), S(y,x)$.  The trivial
    coverage $\C = \set{q}$ is strict, has one factor $\F = \set{q}$,
    and there is an inversion from $(q, x, y)$ to $(q, y, x)$ because
    $S(x,y)$ unifies with $S(y,x)$ (recall that we rename the
    variables before the unification, i.e. the unifier is between
    $R(x)$, $\underline{S(x,y)}$, $S(y,x)$ and its copy $R(x'),
    S(x',y'), \underline{S(y',x')}$).  In the canonical coverage
    $\C^<$ there are three factors, corresponding to $x < y$, $x = y$,
    and $y < x$, and the inversion is between $x < y$ and $y < x$.
\end{example}

\subsection{An Expansion Formula for Coverage}
\label{sec:expansion}

Given a conjunctive query $q$ and a probabilistic structure ${\bf A} =
(A, R^A_1, \ldots, R^A_k)$, we want to compute the probability $p(q)$.
Our main tool is a generalized inclusion-exclusion formula that we
apply to the coverage of a query.

\begin{definition} \label{def:expansion:vars} (Expansion Variables)
  Let $\C = (\F, C)$ be a strict coverage, where $\F = \set{f_1,
    \cdots, f_k}$ is a set of factors and $C$ is a set of subsets of
  $\F$. A set of {\em expansion variables} is a set $\bar{x} =
  \set{\bar{x}_{f_1}, \cdots, \bar{x}_{f_k}}$ such that
\begin{enumerate}
\item $\bar{x}_{f_i} \subseteq Vars(f_i)$ for $1 \leq i \leq k$.
\item If $x \in \bar x_f$ and $x \sqsubset y$, then $y \in \bar x_f$.
\item Any MGU of any two factors $f_i$ and $f_j$ equates an expansion
variable to an expansion variable.
\end{enumerate}
\end{definition}

We use $(\F, C, \bar{x})$ to denote a coverage where we have chosen
the expansion variables. 

\begin{definition} (Unary coverage) A coverage $(\F, C, \bar x)$
is called a {\em unary coverage} if for each $f \in F$, $\bar x_f$
consists of a single variable $r_f$. We call $r_f$ the root variable
in $f$.
\end{definition}
By definition of expansion variables, the root variable must be the
maximal element under $\sqsubset$ order, i.e. must occur in all the
sub-goals of the corresponding factor.

Our first \PTIME\ algorithm (for inversion-free queries) uses a unary
coverage: the discussion in the next few subsections is much easier to
follow if one assumes all coverages to be unary.  Our second \PTIME\ 
algorithm (for queries with erasable inversions) uses a coverage in
which all variables are expansion variables, i.e. $\bar x_f =
Vars(f)$: for that reason our discussion below needs to be more
complex.

For $f \in \F$, let $A_f = A^{|\bar{x}_f|}$, and for $\bar a
\in A_f$, let $f(\bar a)$ denote the query $f[\bar a / \bar{x}_f]$, i.e.,
the conjunctive query obtained by substituting the variables
$\bar{x}_f$ with $\bar a$. The following follows simply from the
definitions:
\begin{eqnarray}
  q & = & \bigvee_{c \in C} \bigwedge_{f \in c} \bigvee_{\bar a \in
  A_f} f(\bar a) 
\label{eq:bigvee} 
\end{eqnarray}

Our next step is to apply the inclusion/exclusion formula to
(\ref{eq:bigvee}). We need some notations. 
We call a subset $\sigma \subseteq \F$ a {\em signature}.
Given $s \subseteq C$, its signature is $sig(s) = \bigcup_{c \in s} c$.

\begin{definition} Given a set $\sigma \subseteq
\F$, define
$$
N(\C, \sigma) = (-1)^{|\sigma|}\sum_{s \subseteq C: sig(s) = \sigma} (-1)^{\mid s \mid} 
$$
\end{definition}

For example, if $C = \set{c_1, c_2, c_3}$ where $c_1 =
\set{f_1,f_2}$,$c_2 = \set{f_2,f_3}$ and $c_3 = \set{f_1,f_3}$, then for 
signature $\sigma = \set{f_1,f_2,f_3}$ we have
$N(\sigma) = (-1)^{|\set{f_1,f_2,f_3}|}(
(-1)^{\mid \set{c_1,c_2}  \mid} +
(-1)^{\mid \set{c_1,c_3}  \mid} +
(-1)^{\mid \set{c_2,c_3}  \mid} +
(-1)^{\mid \set{c_1,c_2,c_3}  \mid}) =
-2$.
\eat{
For the second formula we note
that the complement of $UP(C)$ is $\set{\emptyset, \set{f_1},
\set{f_2}, \set{f_3}}$, hence $N(\sigma) = (-1)^0 + 3 (-1)^1 = -2$.  
}

Given $k$ sets $\bar T = \set{T_{f_1}, \ldots, T_{f_k}}$, where $T_{f_i} \subseteq
A_{f_i}$, we denote its signature $sig(\bar T) = \setof{f}{T_f
\not=\emptyset}$, its cardinality $|\bar T| = \sum_i \mid T_{f_i}|$,
and denote $\F(\bar T)$ the query $\bigwedge_{f \in \F}
\bigwedge_{a \in T_f} f(\bar a)$.  

\begin{definition} (Expansion) Given a coverage $\C$, define its expansion as 
  \begin{eqnarray}
    Exp(\C) & = & \sum_{\bar T} N(\C, sig(\bar T)) (-1)^{\mid \bar T
    \mid} p(\F(\bar T)) \label{eq:n-formula}
  \end{eqnarray}
\end{definition}
We prove the following in the
appendix, using the inclusion/exclusion formula on (\ref{eq:bigvee}):
\begin{theorem} \label{th:n-formula} (Expansion Theorem) If $\C$ is a
coverage for $q$, then $p(q) = Exp(\C)$.
\end{theorem}

Of course, Equation~(\ref{eq:n-formula}) is of exponential size.  To
reduce it, our first goal is to express $p(\F(\bar{T}))$ as the
product $\prod_f \prod_{\bar a \in T_f} p(f(\bar a))$. For that we need to
ensure that any two queries $f(\bar a)$, $\bar a \in A_f$ and $f'(\bar
a')$, $\bar a' \in
A_{f'}$ are independent, and this does not hold in general.  We will
enforce this by restricting the sets $\bar T$ in
Eq.~(\ref{eq:n-formula}) to satisfy some extra conditions, which
we call {\em independence predicates}.  We first illustrate
independence predicates on a running example, then present them in the
general case.  Then we will move to our second goal: finding a
closed form for the sum of products.

\subsection{Running Example}

\label{sec:example}

We give the basic intuition for independence predicates using the
following example.

\begin{example}
\label{ex:independence}
Consider the following query 
$$q = P(x),R(x,y), R(x',y'),S(x')$$ 
and a coverage $\C =
(\F, C, \bar{x})$ where $\F$ consists of the following three queries:
\begin{eqnarray*}
f_1 &=& P(x_1), R(x_1,y_1)\\
f_2 &=& R(x_2,y_2), S(x_2)\\
f_3 &=& P(x_3), R(x_3,y_3), S(x_3)
\end{eqnarray*}
and $C = \set{ \set{f_1, f_2}, \set{f_3}}$ and the expansion variables
are $\bar x_{f_1} = \set{x_1}, \bar x_{f_2} = \set{x_2}, \bar x_{f_3}
= \set{x_3}$. It is easy to verify that $\C$ defined here is indeed a
coverage.  (Here $f_3$ is redundant, i.e. $\set{\set{f_1,f_2}}$ is
already a coverage.  The reason why we include $f_3$ will become clear
later.)  The function $N$ on signatures is as follows: $N(\C,
\set{f_1,f_2}) = 1$, $N(\C, \set{f_3}) = N(\C, \set{f_1, f_2, f_3}) =
-1$ and $N(\C, \sigma) = 0$ for all other $\sigma$.  Thus, the
inclusion-exclusion formula in Theorem~\ref{th:n-formula} gives:
\begin{eqnarray}
p(q) & = & \sum_{\bar T} N(\C, sig(\bar T)) (-1)^{\mid \bar T \mid} p(\F(\bar T))
\label{eq:example1}
\end{eqnarray}
where $\bar T$ is a triplet of sets $\set{T_1,T_2,T_3}$, $|\bar T| =
|T_1| + |T_2| + |T_3|$ and $\F(\bar T) = f_1(T_1)f_2(T_2)f_3(T_3)$.
Consider now three sets $T_1, T_2, T_3$, and let's examine the query
$\F(\bar T)$.  If $T_1 \cap T_2 = T_1 \cap T_3 = T_2 \cap T_3 =
\emptyset$ then $f_i(a)$ is independent from $f_j(a')$, for all $i\neq
j$, or for $i=j$ and $a \neq a'$.  In this case $p(\F(\bar T)$ is a
product $\prod_{i=1,3} \prod_{a \in A} p(f_i(a))$.  We will ensure
that the sets $T_i$ are disjoint in two steps.  First we will show:
\begin{eqnarray}
p(q) & = & \sum_{\bar T \mid T_1 \cap T_2 = \emptyset} N(\C, sig(\bar T)) (-1)^{\mid \bar T \mid}
          p(\F(\bar T)) 
\label{eq:sumind}
\end{eqnarray}
Starting from Eq.(\ref{eq:example1}) we note that $N(\C, sig(\bar T))$
is $\not=0$ for only three signatures, hence $p(q) = p_1 + p_2 + p_3$, where \\\\
\indent \indent \indent \indent $p_1= \sum_{T_1 \neq \emptyset,
  T_2 \neq \emptyset, T_3 = \emptyset} (-1)^{\mid \bar T \mid} p(\F(\bar T))$\\
\indent \indent \indent \indent $p_2=-\sum_{T_1 = T_2 =
  \emptyset, T_3 \neq \emptyset} (-1)^{\mid \bar T \mid} p(\F(\bar T))$\\
\indent \indent \indent \indent $p_3=-\sum_{T_1 \neq \emptyset, T_2
  \neq \emptyset, T_3 \neq \emptyset} (-1)^{\mid \bar T \mid}
p(\F(\bar T))$

Let $p_1^I$ and $p_3^I$ denote the same sums as $p_1$ and $p_3$, but
where $\bar T$ is restricted to satisfy $T_1 \cap T_2 = \emptyset$. To
prove Equation~(\ref{eq:sumind}), all we need is to show is that $p_1
+ p_3 = p_1^I + p_3^I$. In the sum defining $p_3$ denote $T_3' = T_3 -
T_1 \cap T_2$, $T_3'' = T_3 \cap T_1 \cap T_2$ (hence $T_3 = T_3' \cup
T_3''$) and $\bar T' = (T_1, T_2, T_3')$.  We have $p_3 =$

\begin{eqnarray*}
& = & - \hspace{-5mm} \sum_{\scriptsize
\begin{array}{c}
\bar T' \mid T_1 \neq \emptyset, T_2 \neq \emptyset \\
T_3' \cap T_1 \cap T_2 = \emptyset
\end{array}}
\sum_{\scriptsize
\begin{array}{c}
T_3'' \subseteq T_1 \cap T_2 \\
T_3' \cup T_3'' \neq \emptyset
\end{array}} (-1)^{|\bar T|} p(\F(\bar T))  \\
 & = & - \hspace{-5mm} \sum_{\scriptsize
\begin{array}{c}
\bar T' \mid T_1 \neq \emptyset, T_2 \neq \emptyset \\
T_3' \cap T_1 \cap T_2 = \emptyset
\end{array}} (-1)^{|\bar T'|}p(\F(\bar T'))
\hspace{-3mm} \sum_{\scriptsize
\begin{array}{c}
T_3'' \subseteq T_1 \cap T_2 \\
T_3' \cup T_3'' \neq \emptyset
\end{array}} (-1)^{|T_3''|} \\
 & = & p_3^I + 0 + (p_1^I - p_1)
\end{eqnarray*}

The first line simply splits the summation into a sum where
$T_1,T_2,T_3'$ range over subsets of $A$, and an inner sum where
$T_3''$ ranges over subsets of $T_1 \cap T_2$.  The second line holds
because the query $\F(\bar T) = f_1(T_1)f_2(T_2)f_3(T_3')f_3(T_3'')$
is logically equivalent to $f_1(T_1)f_2(T_2)f_3(T_3')$ since $\forall
a \in T_3''$ $f_3(a)$ is $f_1(a)f_2(a)$ and $a$ is in both $T_1$ and
$T_2$.  The last line follows by breaking the sum into three disjoint
sums:
\begin{enumerate}
\item $T_1 \cap T_2 = \emptyset$. Then, $T_3''$ is only allowed to be
the empty set and the inner sum is 1. The total contribution of such
terms is exactly equal to $p_3^I$.
\item $T_1 \cap T_2 \neq \emptyset, T_3' \neq \emptyset$. Then the
  inner sum, $\sum_{T_3''} (-1)^{|T_3''|}$ is 0, because $T_3''$
  ranges over all subsets of $T_1 \cap T_2$.
\item $T_1 \cap T_2 \neq \emptyset, T_3' = \emptyset$. Then the inner
  sum is -1, because $T_3''$ ranges over all subsets of $T_1 \cap T_2$
  except $\emptyset$. The total contribution is $p_1^I - p_1$.
\end{enumerate}
Thus, we have shown Equation~(\ref{eq:sumind}).  Next, we introduce
similar predicates between $T_1, T_3$, and $T_2, T_3$.  This turns out
to be much simpler: we write $T_1$ as $T_1' \cup T_1''$ where $T_1' =
T_1 - T_3$ and $T_1'' = T_1 \cap T_3$. Similarly, we write $T_2$ as
$T_2' \cup T_2''$ with $T_2' = T_2 - T_3$ and $T_2'' = T_2 \cap T_3$.
The query $f_1(T_1)f_2(T_2)f_3(T_3)$ is logically equivalent to
$f_1(T_1')f_2(T_2')f_3(T_3)$ since both $f_1$ and $f_2$ have a mapping
to $f_3$. We now have independence predicates between $T_1'$ and $T_3$
and $T_2'$ and $T_3$. 
\eat{
Also, $(-1)^{|\bar T|}
= (-1)^{|T_1'| + |T_3| + |T_2'| + |T_3| + |T_3|} = (-1)^{|T_1'| +
|T_2'| + |T_3|}$.
}
We replace $\bar T$ with $\bar
T'=(T_1',T_2',T_3',T_1'',T_2'')$. Denoting $\IP(\bar T')$  =  $(T_1'
\cap T_2' = T_1'' \cap T_2'' = T_1' \cap T_3' = T_2' \cap T_3' =
\emptyset, T_1'' \subseteq T_3', T_2'' \subseteq T_3')$, we have:
\begin{eqnarray}
\hspace{-5mm} p(q) & = &  \sum_{\IP(\bar T')} N(\C, sig(\bar T')) (-1)^{\mid \bar T \mid}
          p(\F(\bar T')) \nonumber \\
   & = &  \sum_{\IP(\bar T')} N(\C,
   sig(\bar T')) (-1)^{\mid \bar T \mid} \prod_{i=1,3} \prod_{a \in T_i'}
   p(f_i(a)) \label{eq:sumind2}
\end{eqnarray}
Note that the summation is over five sets $T_1',T_2',T_3',T_1'',T_2''$
but only $T_1',T_2',T_3'$ are used in the compuation of $p$. The
independence predicate $\IP$ allowed us to express $p(\F(\bar T))$ as
a product.  We will show later how to compute this sum.  First, we
need to show how to derive and use independence predicates in general.
\hfill $\Box$
\end{example}

\subsection{Independence Predicates}

Our goal in this section is to define formally independence
predicates.  For unary coverages, an independence predicate is simply
a statement $T_i \cap T_j \neq \emptyset$, but the non-unary case
requires more formalism.  We first introduce a new relational
vocabulary, $\T$ consisting of the relation symbols $T_{f_1}, \cdots,
T_{f_k}$ of arities $|x_{f_1}|, \ldots, |x_{f_k}|$ respectively.  A
structure over this vocabulary is a $k$-tuple of sets $\bar T$; given
a conjunctive query $\phi$ over the vocabulary $\T$, $\bar T \models
\phi$ means that $\phi$ is true on $\bar T$.  For a trivial
illustration, assume $T_{f_1}$, $T_{f_2}$ to be of arity 1, and $\phi
= T_{f_1}(x),T_{f_2}(x)$.  Then $\phi$ states that $T_{f_1} \cap
T_{f_2} \neq \emptyset$.

Suppose we have have two factors $f_i$ and $f_j$ and $\theta$ is any
1-1 substitution on $f_i,f_j$, given in set representation, such that
for all $(x_i,x_j) \in \theta$, $x_i$ is an expansion variable of $f_i$
and $x_j$ is an expansion variable of $f_j$. Define
\begin{eqnarray*}
\theta^R(f_i, f_j) &=& f_i, f_j, \bigwedge_{(x_i, x_j) \in \theta} x_i = x_j\\
\theta^T(f_i, f_j) &=& T_{f_i}(\bar x_{f_i}), T_{f_j}(\bar x_{f_j}), \bigwedge_{(x_i, x_j) \in \theta} x_i = x_j
\end{eqnarray*}
Note that $\theta^R(f_i,f_j)$ is over the vocabulary ${\cal R}$ (same
as the original query $q$), while $\theta^T(f_i, f_j)$ is over the
vocabulary $\T$.  We call them the {\em join query} and the {\em join
  predicate} respectively.  We call the negation of join predicate,
$\texttt{not}(\theta^T(f_i,f_j))$, an {\em independence predicate}.

\begin{example} Consider factors
  $f_1$ and $f_2$ in Example~\ref{ex:independence}, and let $\theta =
  \set{(x_1,x_2)}$. Then, $\theta^R(f_1,f_2) = P(x),R(x,y),S(x)$,
  $\theta^T(f_i,f_j) = T_1(x),T_2(x)$, and the independence predicate
  $\texttt{not}(\theta^T(f_i,f_j))$ says that $T_1$ and $T_2$ are
  disjoint.
\label{ex:independence3}
\end{example}

\eat{
The key property of independence predicates is:

\begin{lemma} \label{lemma:2:independence}
  Suppose $(T_i,T_j) \models \texttt{not}(\theta^T(f_i,f_j))$.  Then
  for all $\bar a \in T_i, \bar a' \in T_j$, $f_i(\bar a)$ is
  independent from $f_j(\bar a')$.
\end{lemma}
}

The key property of independence predicates is the following:  If
$T_i, T_j$ satisfy all independence predicates between $f_i$ and
$f_j$, then for all $\bar a \in T_i$ and $\bar a' \in T_j$, $f_i(\bar
a)$ and $f_j(\bar a')$ are independent.

\subsection{Hierarchical Closure}
\label{sec:hierarchical}

Recall from Example~\ref{ex:independence} that, in order to introduce
an independence predicate between two sets $T_1, T_2$ we needed to use
the join query of their factors, $f_3(x) = f_1(x),f_2(x)$.  In
general, the join query between two factors in $\F$ is not necessarily
in $\F$ ($f_3$ was redundant in Example~\ref{ex:independence}).  Thus,
we will proceed as follows.  Starting from a coverage $\C$ we will add
join queries repeatedly until we obtain its {\em hierarchical
  closure}, denoted $\C^*$, then we will introduce independence
predicates.  Computing $\C^*$ is straightforward when $\C$ is an
inversion-free coverage (which is the case for our first \PTIME\ 
algorithm), but when $\C$ has inversions then some join queries are
non-hierarchical and we cannot add them to $\C^*$.  We define next
$\C^*$ in the general case.
Let $\C = (\F, C, \bar{x})$ be any coverage with a set of expansion
variables $\bar x$.

\begin{definition} \em Given two factors $f_1$ and $f_2$, with
expansion variables $\bar x_{f_1}$ and $\bar x_{f_2}$, and a $MGU$
given by the set representation $\theta$, the {\em hierarchical
unifer} $\theta_u$ is the maximal subset of $\theta$ such that:
\begin{enumerate}
\item $(x,y) \in \theta_u \Rightarrow x \in \bar x_{f_1}, y \in \bar x_{f_2}$
\item If $(x,y) \in \theta_u$ and $(x',y')$ is such that $x \sqsubseteq x'$
or $y \sqsubseteq y'$ and $(x',y') \in \theta$, then $(x',y') \in \theta_u$. 
\item The query $\theta_u^R(f_1,f_2)$ is hierarchical.
\end{enumerate}

\eat{
Since we want to add a predicates on $T$, we can only
refer to the expansion variables of the factors. The first condition
ensures this. The third condition ensures that we only generate
hierarchical queries.
}

It can be shown that $\theta_u$ is uniquely determined. If $\theta_u$
is non-empty, we say that $f_1$ and $f_2$ can be {\em hierarchical
  joined} using $\theta$ and call the query $\theta_u^R(f_1,f_2)$ the
{\em hierarchical join} of $f_1$ and $f_2$, and $\theta_u^T(f_1,f_2)$
the {\em hierarchical join predicate}.
\end{definition}

\begin{example} \label{ex:hunif}
Let
\begin{eqnarray*}
f_1 =&R(r,x), & S(r,x,y), U(a,r), U(r,z), V(r,z)\\
f_2 =&        & S(r',x',y'),T(r',y'),V(a,r')
\end{eqnarray*}
and $\theta = \set{(r,r'),(x,x'),(y,y')}$ be the MGU of the two $S$
sub-goals. Then, the hierarchical unifier is $\theta_u = \{(r,r')\}$.
If we include any of $(x,x')$ or $(y,y')$, we will have to include the
other because $x \sqsubset y$ and $x' \sqsupset y'$, and then the join
will not be hierarchical.  The hierarchical join for this unifier is
\begin{eqnarray*}
\theta_u^R(f_1,f_2) = & R(r,x), & S(r,x,y), U(a,r), U(r,z), V(r,z)\\
        && S(r,x',y'),T(r,y'),V(a,r)
\end{eqnarray*}
and the set of expansion variables of the join is $\set{r}$.
\hfill $\Box$.
\end{example}
Starting from the factors $\F$, we construct a set $\H$, a function
\factors\ from $\H$ to subsets of $\F$, and a set of expansion
variables $\bar x_h$ for $h \in \H$.  This is done inductively as
follows:
\begin{enumerate}
\item For each $f \in \F$, add $f$ to $\H$ and let
$\factors(f) = \{f\}$.
\item For any two queries $h_1$, $h_2$ in $\H$, and any MGU $\theta$
  between $h_1$ and $h_2$, let $h = \theta^R_u(h_1,h_2)$ be their
  hierarchical join.  Then add $h$ to $\H$, define $\factors(h) =
  \factors(h_1) \cup \factors(h_2)$; define $\bar x_h = \theta_u(\bar
  x_{h_1} \cup \bar x_{h_2})$.
\end{enumerate}

We need to show that $\H$ is finite.  This follows from:

\begin{lemma} \label{lemma:finite-hierarchical}
  Given a fixed relational vocabulary $\cal R$ and a fixed set of
  constants $C$, the number of distinct hierarchical queries over
  $\cal R$ and $C$ is finite.
\end{lemma}

Define $\F^*$ to be the subset of $\H$ containing queries that are
either inversion-free or in $\F$.

\begin{definition} (Hierarchical Closure) Given a coverage $\C = (\F,
  C, \bar{x})$, its {\em hierarchical closure} is $\C^* = (\F^*, C^*,
  \bar{x}^*)$ where $\F^*$, $\bar{x}^*$ are defined above and:
  $$C^* = \setof{c}{c \subseteq \F^*, \bigcup_{f \in c}
    \factors(f) \in C}$$
\end{definition}
Note that $\C^*$ is indeed a coverage since the set $\F^*$
contains the set $\F$, the set $C^*$ contains the set $C$, and the
expansion variables satisfy the conditions in
Def.~\ref{def:expansion:vars}.
Let $\IP(\C^*)$ be the conjunction of $\texttt{not}(jp)$, where $jp$
ranges over all possible hierarchical join predicates in $\F^*$.

\begin{lemma} If $T \models \IP(\C^*)$, then 
$$p(\F(q)) = \prod_{f \in \F^*} \prod_{a \in T_f} p(f(\bar a))$$
\label{lemma:independence}
\end{lemma}

Finally, we look at conditions under which we can add the predicate
$\IP(\C^*)$ over $\bar T$. We divide the join predicates into two
disjoint sets, {\em trivial} and {\em non-trivial}. A join predicate
between factors $h_i$ and $h_j$ is called trivial if the join query is
equivalent to either $h_i$ or $h_j$, and is called non-trivial
otherwise. We write $\IP(\C^*)$ as $\IP^n(\C^*) \wedge \IP^t(\C^*)$,
where $\IP^n(\C^*)$ is the conjunction of $\texttt{not}(jp)$ over all
non-trivial join predicates $jp$, and $\IP^t(\C^*)$ is the conjunction
over all trivial join predicates.

\begin{definition} \label{def:eraser} (Eraser) Given a hierarchical
  join $jq = 
  \theta^R_u(f_i,f_j)$, an {\em eraser} for $jp$ is a set of factors
  $E \subseteq \F$ s.t.:
   \begin{enumerate}
       \item $\forall q \in E$, there is a homomorphism from $q$ to
       $jq$. 
     \item $\forall \sigma \subseteq \F$, $N(\C, \sigma \cup \{f_i,
       f_j\}) = N(\C, \sigma \cup \{f_i,f_j\} \cup E)$.
    \end{enumerate}
\end{definition}

\begin{theorem} Let $q$ be a query such that  every hierarchical
  join query $jq = \theta^R_u(f_i,f_j)$ between two factors in $\F^*$
  has an eraser.  Then,
\begin{eqnarray*}
p(q) &=& \sum_{\bar T \mid T \models \IP^n(\C^*)} N(\C^*, sig(\bar T))
        (-1)^{\mid \bar T \mid} p(\F(\bar T))
\end{eqnarray*}
\label{th:independence}
\end{theorem}

The theorem allows us to add all possible non-trivial independence
predicates over the summation.  If the hierarchical join query $jp$ is
inversion-free, then it belongs to $\F^*$ and it is its own eraser
(i.e. $E=\set{jp}$ satisfies both conditions above).  We can use it to
separate $T_i$ from $T_j$.  In particular if $q$ is inversion-free,
then any hierarchical join query has an eraser, and all sets can be
separated.  But if $jp$ has an inversion, then $jp$ does not belong to
$\F^*$ and we must find some different query (queries) in $\F^*$ that
can be used to separate $T_i$ from $T_j$.

\eat{ The proof (in the appendix) has two steps, following the
  intuition in Example~\ref{ex:independence}.  First it separates
  $T_i, T_j$ by using their eraser $T_k$, then separate each of them
  from their eraser.  }

\begin{example} 
\label{ex:independence4}
Let's revisit the query in Example~\ref{ex:independence}.  We had
$q = P(x), R(x,y),R(x',y'),S(x')$.
Suppose we start from
the trivial coverage $\C_0 = \set{q}$, with two factors $\F_0 =
\set{f_1, f_2}$ (see notations in Example~\ref{ex:independence}), and
suppose we chose a single expansion variable for $f_1$ and $f_2$,
namely $x_1, x_2$ respectively.  Its hierarchical closure adds the
join query $f_3$ between $f_1$ and $f_2$. The coverage $\C^*_0$
contains the following covers: $\set{f_1,f_2}$, $\set{f_3}$,
$\set{f_1,f_3}$, $\set{f_2,f_3}$ and $\set{f_1,f_2,f_3}$.
\eat{
is exactly the
coverage $\C$ that we used in Example~\ref{ex:independence}.  Once we
have closed the coverage under join predicates, we apply
Theorem~\ref{th:independence}: this gives us precisely Eq.
(\ref{eq:sumind2}) that we derived in Example~\ref{ex:independence}.
}
\eat{
 It has two factors $f_1 = P(x),R(x,y)$ and
$f_2 = R(x',y'),S(x')$. Thus $\F = \set{f_1,f_2}$ and $C =
\set{\set{f_1,f_2}}$. Consider expansion variables $\bar x$ given by
$\bar x_{f_1} = x$ and $\bar x_{f_2} = y$.

The set $\H$ consists of $f_1$, $f_2$ and their hierarchical unifier
$f_{3}$ which simply equates their root variables: $f_{3} = P(x_1),
R(x_1,y_1), R(x_1, y_1'), S(x_1)$. The expansion variable of $f_{3}$
consists of $\set{x_1}$. None of the three queries have inversion, so
$\F^* = \H$. 

Let $T_1,T_2,T_{3}$ denote the sets corresponding to the three
unifiers in the expansion. There are three join predicates: (1)
between $f_1$ and $f_2$ given by $jp_1 = T_1(r)T_2(r)$, (2) between
$f_1$ and $f_{3}$ given by $jp_2 = T_1(r)T_{3}(r)$ and (3) between
$f_2$ and $f_{3}$ given by $jp_3 = T_{2}(r)T_{3}(r)$.  Let $\IP =
\texttt{not}(jp_1) \wedge \texttt{not}(jp_2) \wedge
\texttt{not}(jp_3)$. Thus, $\IP$ is equivalent to the condition that
$T_1,T_2,T_{3}$ are disjoint. Using Theorem~\ref{th:independence}, we
get
\begin{eqnarray}
\label{eq:example4}
p(q) = \sum_{\bar T \mid T_1,T_2,T_3~\mbox{disjoint}} N(\C^*,\bar T) (-1)^{|\bar T|}
p(\F(\bar T))
\end{eqnarray}
}
\end{example}

Thus, we have expressed the probability of a query $p(q)$ using the
sum in Theorem~\ref{th:independence}.  This is still exponential in
size, and now we will show how compute a closed form for that sum.
Here we will use different techniques for the two \PTIME\ algorithm.
In the first algorithm (for inversion-free queries) the coverage is
unary, and all independence predicates are of the form $T_i \cap T_j
\neq 0$: here we derive closed forms directly.  In the second
algorithm (for queries with erasable inversions) the independence
predicates are more complex: in this case we will reduce the sum to
the probability of an inversion-free query $\phi$ over the $\T$
vocabulary, thus bootstrapping the first \PTIME\ algorithm.

\section{PTIME Algorithms}
\label{sec:ptime}

In this section, we establish one-half of the dichotomy by proving
Theorem~\ref{th:dichotomy}(2). We start by computing simple sums over
functions on sets, then use it to give a \PTIME\ algorithm for queries
without inversion and finally give the general \PTIME\ algorithm for
queries that have erasers for all inversions.

\subsection{Simple Sums}
\label{sec:summations}

Let $A = \set{1, \ldots, N}$, $\bar g = (g_1, \ldots, g_k)$ be $k$
functions $g_i : A \rightarrow {\bf R}$, $i=1,\ldots, k$, and $\bar T
= (T_1, \ldots, T_k)$ a $k$-tuple of subsets of $A$.  Denote $\bar
g(\bar T) = g_1(T_1) \cdots g_k(T_k)$, where $g_i(T_i) = \prod_{\bar a
  \in T_i} g_i(\bar a)$.  $\emptyset \not= \bar T$ abbreviates
$\emptyset \not=T_1, \ldots, \emptyset \not=T_k$.  Let $\phi$ be a
conjunction of statements of the form $T_i \cap T_j = \emptyset$ or
$T_i \subseteq T_j$, and define:
$S_\phi = \setof{\sigma}{\sigma \subseteq [k], \forall i, j \in
  \sigma, \phi \not\models T_i \cap T_j =\emptyset}$ $\cap$
  $\setof{\sigma}{\sigma \subseteq [k], \forall i \in
  \sigma, j \not\in \sigma,  \phi \not\models T_i \subseteq T_j}$.

\eat{
set of signatures (i.e. subsets of $[k]$) and denote
$\bar T \models \Phi$ the statement $\setof{i}{T_i \not= \emptyset}
\in \Phi$.  For a signature $\sigma \subseteq [k]$ we denote $\bar
g_\sigma$ the family of functions $(g_i)_{i \in \sigma}$.
}

\begin{definition}  Denote the following sums:
  \begin{eqnarray*}
    \summation_\phi \bar g & = & \sum_{\bar T \subseteq  A, \phi} \bar g(\bar T) \\
    \summation^+_\phi \bar g & = & \sum_{\emptyset \not= \bar T \subseteq  A, \phi}  \bar g(\bar T) 
  \end{eqnarray*}
\end{definition}

For $\sigma \subseteq [k]$, denote $\bar g_\sigma$ the family of
functions $(g_i)_{i \in \sigma}$.

  \begin{proposition} \label{prop:sums}
The following closed forms hold:
    \begin{eqnarray*}
      \summation_\phi \bar g & = & \prod_{a \in A} \sum_{\sigma \in S_\phi} \prod_{i
      \in \sigma} g_i(\bar a) \\
      \summation^+_\phi \bar g & = & \sum_{\sigma \subseteq [k]} 
      (-1)^{k -\mid \sigma \mid} \summation \bar  g_\sigma 
    \end{eqnarray*}
    Moreover, the expressions above have sizes 
    $O(k2^kN)$ and $O(k2^{2k}N)$ respectively, hence all have an
    expression size that is linear in $N$.
  \end{proposition}

\begin{example} \label{ex:sum}
Consider four functions $g_i : A \rightarrow R$, $i=1,2,3,4$, and suppose
we want to compute the following sum:
$$
\sum_{T_1 \cap T_2 = \emptyset, T_2 \cap T_3
= \emptyset, T_4 \subseteq T_2} g_1(T_1) g_2(T_2) g_3(T_3) g_4(T_4)
$$
In our notation, this is $\summation_\phi \bar g$, where $\phi$ is
$T_1 \cap T_2 = \emptyset \wedge T_2 \cap T_3 = \emptyset \wedge T_4
\subseteq T_2$. The set
$S_\phi$ is $\set{\emptyset, \set{1}, \set{2}, \set{2,4}, \set{3}, \set{1,3}}$.
Thus, the expression for the sum is
$$\prod_{a \in A} (1+g_1(a)+g_2(a) + g_2(a)g_4(a) + g_3(a) + g_1(a)g_3(a))$$ 
The size of this expression is $8N$, where $N$ is the size of $A$.
\eat{
\begin{eqnarray*}
\summation (g_i)  & = & (1+g_i(1))(1+g_i(2)) \cdots  (1+g_i(N)) \\
\summation^+ (g_i)  & = & \summation (g_i) - 1 \\
\summation(g_1,g_2) & =
& (1+g_1(1)+g_2(1))\cdots(1+g_1(N)+g_2(N)) \\
\summation^+(g_1,g_2) & = & \summation(g_1,g_2) - \summation(g_1) -
\summation(g_2) + 1 \\
\summation_\Phi(g_1,g_2) & = & \summation^+(g_1,g_2) + \summation^+(g_1) + \summation^+(g_2)
\end{eqnarray*}
}
\end{example}

\subsection{PTIME for Inversion-Free Queries}
\label{sec:noinversions-ptime}

Let $q$ be an inversion-free query. We give now a \PTIME\ 
algorithm for computing $q$ on a probabilistic structure.

\begin{theorem} If $q$ has no inversions then
  $q$ has a unary coverage.
\label{th:roots}
\end{theorem}

This says that we can choose for each factor $f$ a single {\em root}
variable $r_f$ s.t. any MGU between two (not necessarily distinct)
factors $f, f'$ maps $r_f$ to $r_{f'}$: the proof in the Appendix uses
the canonical coverage $\C^<$, considers for each factor $f$ all
maximal variables under $\sqsupseteq$, and chooses as root variable
the maximum variable under $>$.  Note that for queries with inversions
Theorem~\ref{th:roots} fails (recall the queries $H_k$).

\begin{example}
We illustrate Theorem~\ref{th:roots} on two queries.
\begin{eqnarray*}
q_1 & = & R(x,y),S(x,y),S(x',y'),T(y') \\
q_2 & = & R(x,y),R(y,x)
\end{eqnarray*}
In the trivial coverage $\C = \set{q_1}$ for $q_1$ the factors are
\begin{eqnarray*}
f_1 = R(x,y),S(x,y) & & f_2 = S(x',y'),T(y')
\end{eqnarray*}
We see that $r_{f_1} = \set{y}$ and $r_{f_2} = \set{y'}$ satisfy the
properties of Theorem~\ref{th:roots} (there are two maximal variables
for $f_1$, but we have to pick $y$ because it unifies with $y'$).  For
$q_2$, the trivial coverage $\C = \set{q_3}$ does not work since there
is a unifier that unifiers $x$ with $y$, and exactly one of them can
be the expansion variable.  On the other hand, consider the following
coverage:
\begin{eqnarray*}
f_1 = R(x_1,y_1),R(y_1,x_1),x_1 > y_1 & &
f_2 = R(x,x)
\end{eqnarray*}
now we can set $r_{f_1} = x_1$ and $r_{f_2} = x$.
\hfill $\Box$
\end{example}

Now, let $q$ be a query without inversion and $\C = (\F, C, \bar x)$
be any unary coverage.  Let $\C^* = (\F^*, C^*, \bar x^*)$ be the
hierarchical closure of $\C$.  Theorem~\ref{th:independence} applied
to this unary coverage gives:
$$p(q) = \sum_{\bar T \mid T \models \IP^n(\C^*)} N(\C^*, sig(\bar T))
(-1)^{\mid \bar T \mid} p(\F^*(\bar T))
p(f(\bar a))$$
All the sets in $T$ have arity 1, since $\C^*$ is also unary, hence
each join predicate has the form $T_i(x),T_j(x)$ which is equivalent
to $T_i \cap T_j \neq \emptyset$, hence $\IP(\C^*)$ is a conjunction
of predicates of the form $T_i \cap T_j = \emptyset$.

So far we have only added the independence predicates $\IP^n(\C^*)$,
i.e. independence predicates between those pairs $h_i$ and $h_j$
for which the join query is not equivalent to either $h_i$
and $h_j$. Next, we add independence predicates between the remaining
pairs. We generalize our technique of Example~\ref{ex:independence}.
We replace $\bar T$ with $\bar T'$, where $\bar T'$ contains all the
sets in $\bar T$ along with some additional sets. For each $h_i, h_j$
such that their hierarchical join is equivalent to $h_j$, $\bar T'$
contains an additional set $T_{i,j}$. Denote $\IP^l(\C^*)$ the conjunction of
the following predicates 
\begin{itemize}
\item A predicate $T_{i,j} \subseteq T_j$ for all $T_{i,j}$, $T_j$ in
$\bar T'$
\item A predicate $T_{i_1,j} \cap T_{i_2,j} = \emptyset$ for all
$T_{i_1,j},T_{i_2,j}$ in $\bar T'$ such that there is a predicate
$T_{i_1} \cap T_{i_2} = \emptyset$ in $\IP(\C^*)$.
\end{itemize}
Let $\IP'(\C^*)$ denote the conjunction of $\IP(\C^*)$ and $\IP^l(\C^*)$.
Then, we obtain $p(q) =$
\begin{eqnarray*}
\sum_{\bar T \mid T \models \IP'(\C^*) \wedge \IP^l(\C^*)}
N(\C^*, sig(\bar T)) (-1)^{\mid \bar T \mid} \prod_{f
\in \F^*} \prod_{\bar a \in T_f} p(f(\bar a))
\end{eqnarray*}
Corresponding to each $T_i \in \bar T'$, let $g_{i} : A \rightarrow {\bf R}$
denote the function $g_{i}(\bar a) = -p(f_i(\bar a))$. Also,
corresponding to each $T_{i,j} \in \bar T'$, let $g_{i,j}$ denote the
function $g_{i,j}(\bar a) = 1$.

\begin{theorem} \label{th:inversion-free-prob-recur} Let $q$ be inversion-free.
\begin{enumerate}
\item The probability of $q$ is given by
 \begin{eqnarray}
    p(q) & = & \sum_{\sigma \subseteq \F^*} N(\C^*, \sigma)
    \summation^+_{\IP(\C^*) \wedge \IP^l(\C^*)} \bar g_\sigma
  \label{eq:prob-recur}
  \end{eqnarray}
where $\summation^+$ ranges over all sets of the form $\bar T'$.
\item For each $f \in \F^*$, $f(\bar a)$ is an inversion-free query.
\end{enumerate}
\end{theorem}

We use Proposition~\ref{prop:sums} to write a closed-form expression
for Equation~(\ref{eq:prob-recur}) in terms of the probabilities
$g_f(\bar a) = p(f(\bar a))$ for $f \in \F^*$. Since each of these queries is
inversion-free, we recursively apply Equation~(\ref{eq:prob-recur}) to
compute their probabilities.  For any query $q$, let $V(q)$ denote the
maximum number of distinct variables in any single sub-goal of $q$.
Clearly, for any factor $f$, $V(f(\bar a)) < V(f) \leq V(q)$ (since
$\bar a$ substitutes a variable in every sub-goal). Thus, the depth of
the recursion is bounded by $V(q)$.

\begin{corollary} If $q$ is an inversion-free query, then $p(q)$ can
  be expressed as a formula of size $O(N^{V(q)})$, where $N$ is the
  size of the domain.  In particular $q$ is in \PTIME.
\end{corollary}

\begin{example} \label{ex:no-inversion-ptime} Continuing our running
  example from Example~\ref{ex:independence}, recall that $p(q)$ is
  given by Equation~(\ref{eq:sumind2}). Let $\bar T' =
  (T_1,T_2,T_3,T_{1,3}, T_{2,3})$. Denoting $g_i(a) =
  -p(f_i(a))$, for $i=1,2,3$, $\phi \equiv (T_1 \cap T_2 = \emptyset)$
  and $\psi \equiv (T_1 \cap T_2 = \emptyset) \wedge (T_1 \cap T_3 =
  \emptyset) \wedge (T_2 \cap T_3 = \emptyset) \wedge (T_{1,3} \cap
  T_{2,3} = \emptyset) \wedge (T_{1,3} \subseteq T_3) \wedge (T_{2,3}
  \subseteq T_3)$:
  $$p(q) = \summation^+_\phi(g_1,g_2) + \summation^+(g_{3}) +
  \summation^+_\psi(g_1,g_2,g_{3})$$
  Now apply Prop.~\ref{prop:sums}
  to each expression, e.g. $\summation^+_\psi(g_1,g_2,g_3) =
  \summation_\psi(g_1,g_2,g_3) - \summation_\psi(g_1,g_2) - \ldots$
  Each sum in turn has a closed form. 
  \eat{, e.g.
  $\summation_\psi(g_1,g_2,g_3) = \prod_{a \in
    A}(1+g_1(a)+g_2(a)+g_3(a))$.
    }Furthermore, each $f_i(a)$ is a
  query with a single variable ($y$ or $y'$), hence each $g_i(a) =
  p(f_i(a))$ can be computed inductively.
\end{example}

Appendix~\ref{app:examples} gives example of inversion-free queries,
showing several subtleties that were left out from the text.

\eat{
\subsection{Discussion}

{\bf No self-joins} We have shown in Sec.~\ref{subsec:overview} a very
simple formula (\ref{eq:safeplan}) computing $p(q)$ for the case when
$q$ is a hierarchical query without selfjoin.  In our previous
work~\cite{dalvi04prob} we have shown how that formula can be executed
effectively inside a relational database engine, and more recently
have implemented it in MystiQ.  It turns out that if one applies the
general formula (\ref{eq:prob-recur}) to the special case of queries
without self-joins then one recovers the simpler formula
(\ref{eq:safeplan}).  Thus suggest that the more general formula could
be pushed effectively in the database engine in more general cases.

{\bf Query Complexity} When unfolded recursively the formula
(\ref{eq:prob-recur}) is exponential in the size of the query.  This
should be contrasted with the deterministic case: for a fixed
vocabulary where all relations have arity $\leq a$, and every
hierarchical query has tree width $\leq a$, hence its query complexity
is in PTIME.  A related question is the complexity of checking whether
a query is inversion-free: as a preliminary result we have shown that
checking if a given coverage has an inversion is NP-complete.  A study
of the query complexity on probabilistic structures is beyond the
scope of this paper.

}

{\bf Queries with Negated Subgoals} The \PTIME\ algorithm in this
section can be extended to queries with negated sub-goals.  

\begin{definition}
  A {\em conjunctive query with negations} is a query $q = \exists
  \bar x.(\varphi_1 \wedge \ldots \wedge \varphi_k)$, where each
  $\varphi_i$ is either a positive sub-goal $R(t)$, or a negative
  sub-goal $\texttt{not}(R(t))$, or an arithmetic predicate. The query
  $q$ is said to be inversion-free if the conjunctive query obtained
  by replacing each $\texttt{not}(R(t))$ sub-goal with $R(t)$ sub-goal
  is inversion-free.
\end{definition}

\begin{definition} (Inversion-free property) A property $\phi$ is
called {\em inversion-free property} if it can be expressed as a
Boolean combination of queries $\set{q_1, \cdots, q_m}$ such that each
$q_i$ is a conjunctive query with negation and the query $q_1q_2
\cdots q_m$ is inversion-free. 
\end{definition}

\begin{theorem} Let $\phi$ be any inversion-free property.
Then, computing $p(\phi)$ is in \PTIME.
\end{theorem}

\begin{proof} (Sketch)
Consider a single inversion-free conjunctive query with negation.
The same recurrence formula in Theorem~\ref{th:inversion-free-prob-recur} 
applies, the only difference is during recursion we will reach negated
constant sub goals: $p(\texttt{not}(R(a,b,c)))$ is simply
$1-p(R(a,b,c))$.  For any general $\phi$, use inclusion/exclusion
formula to reduce it to conjunctive queries with
negations, each of which is inversion-free.
\end{proof}

\subsection{Complex Sums}

In Section~\ref{sec:noinversions-ptime}, we used simple sums to give a
\PTIME\ algorithm for inversion-free queries. Here, we show that the
\PTIME\ algorithm can be used to compute closed formulas for
complex sums. We call this the {\em bootstrapping technique}.

{\bf Bootstrapping:} Let $\bar g = (g_1, \ldots, g_k)$ be a family of
functions, $g_i : A^{r_i} \rightarrow {\bf R}$, where the arity of
$g_i$ is $r_i$.  We want to compute sums of the form $\texttt{sum} =
\sum_{\bar S \mid \phi} \bar g(\bar S)$, where $\phi$ is a complex
predicate.  We cannot use the summations of
Section~\ref{sec:summations}, which only apply when $g_i$ are unary.  Instead, we use a bootstrapping
technique to reduce this problem back to evaluating an inversion-free
query on a probabilistic database, and use the \PTIME\ algorithm of
Section~\ref{sec:noinversions-ptime}.  The basic principle is that we
can reduce the problem to the evaluation of $\phi$ over a
probabilistic database. Create an probabilistic instance of
${\cal S}$, where, assuming $k=1$ for simplicity, for each tuple $\bar
\bar a \in S$, set its probability to $p(\bar a) = g(\bar a)/(1 + g(\bar a))$.
Then, the probability of $\phi$ over this instance is $p(\phi) =
\sum_{S} \prod_{\bar a \in S} p(\bar a) \prod_{\bar a \not\in S} (1 -
p(\bar a)) = \prod_{\bar a} 1/(1+g(\bar a)) \sum_{S \mid \phi} g(S) =
\prod_{\bar a}
1/(1+g(\bar a)) \texttt{sum}$.  Thus, we can compute \texttt{sum} in \PTIME\ if
we can evaluate the query $\phi$ in \PTIME.

\begin{theorem}
  Let $\phi$ be an inversion-free property. Then $\sum_{\bar S \mid
  \phi} \bar g(\bar S)$ has a closed form polynomial in domain size.
\label{th:bootstraping}
\end{theorem}

\subsection{The General PTIME Algorithm}
\label{sec:ptime-general}

Let $q$ be a conjunctive query and let $\C = (\F, C)$ be a strict
coverage for $q$ and let $\H$ be the set of hierarchical unifiers, as
defined in Section~\ref{sec:hierarchical}. Suppose the following
holds: for every hierarchical join predicate $jp=\theta^T(h_i,h_j)$
between two factors in $\H$, the join query $jq = \theta^R(f_i,f_j)$ 
has an eraser. We will show here that $q$ is in \PTIME, thus
proving Theorem~\ref{th:dichotomy}(2).

We set the expansion variables $\bar x$ to include all variables, i.e.
$\bar x_f = Vars(f)$ for all $f \in \F$. Let $\C^* = (\F^*, C^*,
\bar{x}^*)$ be the hierarchical closure of $\C$. By
Theorem~\ref{th:independence}, we have $p(q) = Exp(\C^*)$, where
\begin{eqnarray}
Exp(\C^*) = \sum_{\bar T \mid \IP^n(\C^*)} N(\C^*, sig(\bar T))
             (-1)^{\mid \bar T \mid} p(\F^*(\bar T)) \nonumber\\
          = \sum_\sigma N(\C^*, \sigma) \sum_{\bar T \mid \IP^n(\C^*), sig(\bar T) =
            \sigma}  (-1)^{\mid \bar T \mid} p(\F^*(\bar T)) \label{eq:exp1}
\end{eqnarray}
Before we proceed, we illustrate with an example:
\begin{example}
\label{ex:eraserquery2}
Consider the query $q$ in Example~\ref{ex:eraserquery}
Although $q$ has an inversion (between the two $S$ Subgoals) we have
argued in Sec.~\ref{subsec:overview} that it is in \PTIME.
Importantly, the third line of constants sub goals plays a critical
role: if we removed it, the query becomes \#\P-hard.

\noindent Consider the coverage $\C = (\F, C, \bar x)$, where $\F$
is\footnote{Strictly speaking each constant sub-goal $R(a)$,
  $S(a,b,c)$, $U(a,a)$ should be a distinct factor.}:
\begin{eqnarray*}  
f_1 &=& R(r,x),S(r,x,y),U(a,r),U(r,z),V(r,z), r \neq a\\
f_2 &=& S(r',x',y'),T(r',y'),V(a,r'), r' \neq a\\
f_3 &=& U(a,z'),V(a,z')\\
f_4 &=& R(a), S(a,b,c), U(a,a)
\end{eqnarray*}
and $C = \set{ \set{f_1, f_2, f_4}, \set{f_2,f_3,f_4}}$. We cannot
simply take the root variables $r$, $r'$, and $z'$ as expansion
variables and proceed with the recurrence formula in
Th.~\ref{th:inversion-free-prob-recur}, because the query $f_{12} =
f_1(r)f_2(r)$ is \#\P-hard.  We must keep all variables as expansion
variables to avoid the inversion.  Thus, the root unifiers $\H$ are
(recall Example~\ref{ex:hunif}):
\begin{eqnarray*}
f_{12} &=&  f_1, f_2, r = r'\\
f_{23} &=&  f_2, f_3, r' = z'\\
f_{13} &=&  f_1, f_3, r  = z'\\
f_{123} &=& f_1, f_2, f_3, r = r' = z'
\end{eqnarray*}
Out of these, $f_{12}$ and $f_{123}$ have inversions, thus $\F^*(q) =
\set{f_1, f_2, f_3, f_4, f_{23}, f_{13}}$. In the expansion
$Exp(\C^*)$, there are sets $T_1,T_2,T_3,T_4,T_{23},T_{13}$ but note
that they are not unary, e.g. $T_1$ has arity 4 as $\bar x_{f_1} =
\set{r,x,y,z}$.  The critical question is how to separate now $T_1$
from $T_2$, since we don't have the factor $f_{12}$.  Here we use the
fact that there exists a homomorphism $f_3 \rightarrow f_{12}$, thus
$f_3$ is an eraser between $f_1$ and $f_2$ and will use $f_3$ to
separate $T_1$, $T_2$.  The definition of an eraser
(Def.~\ref{def:eraser}) requires us to check $\forall \sigma$, $N(\C,
\sigma \cup \{f_1, f_2\}) = N(\C, \sigma \cup \{f_1,f_2,f_3\})$.  The
only $\sigma$ that makes both $N$'s non-zero is $\set{f_4}$ (and
supersets), and indeed the two numbers are equal to $+1$.  It is
interesting to note that, if we delete the last line from $q$, then we
have the same set of factors but a new coverage
$C'=\set{\set{f_1,f_2},\set{f_2,f_3,f_4}}$: then $f_3$ is no longer an
eraser because for $\sigma=\emptyset$ we have $N(\set{f_1,f_2})=1$ and
$N(\set{f_1,f_2,f_3})=0$.  Continuing the example, we conclude that,
with aid from the eraser, we can now insert all independence predicates.
We have to keep in mind, however, that these predicates are no longer
simple disjointness conditions e.g. the predicate between $T_1$ and $T_2$ is
the negation of the query $T_1(r,x,y,z), T_2(r,x',y')$.  \hfill $\Box$
\end{example}

We now focus on each of the inner sums in Equation~(\ref{eq:exp1}).
\eat{
which is of the form

\begin{equation}
\sum_{\bar T \mid \IP(\C^*), sig(\bar T) = \sigma}  (-1)^{\mid \bar T
\mid} p(\F^*(\bar T))
\label{eq:exp2}
\end{equation}
}
We want to reduce it to evaluation of an inversion-free
property, but there are two problems. First, the predicate $\IP^n(\C^*)$
over $\bar T$ is not an inversion-free property. Second, we still need
to add the predicates $\IP^t(\C^*)$ to make $p(\F^*(\bar T))$
multiplicative. To solve these problems, we apply a preprocessing step on
Equation~{\ref{eq:exp1}, which we call the {\em change of basis}. In
this step, we group ${\bar T}$ that generate the same $\F^*(\bar T)$
and sum over these groups. 

\begin{example}
\label{ex:changebasis}
Consider a factor $f = R_1(x,y),R_2(y,z)$. We look at the set
$T(x,y,z)$ corresponding to this factor, which is a ternary set since
$\bar x_f = \set{x,y,z}$. For every $T$, define $S^0 = \pi_y(T)$, $S^1
= \pi_{xy}(T)$ and $S^2 = \pi_y(T)$, hence $T = S^0 \Join S^1 \Join
S^2$ (natural join).  Clearly, $S^0,S^1,S^2$ satisfy the predicate
$S^0=\pi_y(S^1)=\pi_y(S^2)$. Consider the sum
\begin{equation}
\sum_{\bar T} (-1)^{|\bar T|} p(f(\bar T)) \label{eq:changebasis-example}
\end{equation}
We group all $T$ that generate the same $S^0,S^1,S^2$ and show
that the summation in Eq.~\ref{eq:changebasis-example} is
equivalent to the following:
$$
\sum_{\scriptsize
\begin{array}{c}
{S^1,S^2, S^0} \mid\\
S^0 = \pi_y(S^1) =\pi_y(S^2)
\end{array}} (-1)^{|S^1| + |S^2| +
|S^0|} p(R_1(S^1)R_2(S^2)) 
$$
Thus, we have changed the basis of summation from $T$ to
$S^0,S^1,S^2$.
\hfill $\Box$
\end{example}
The change of basis introduces some new predicates between sets, which
we call the {\em link predicates}, e.g.  predicates of the form $S^0 =
\pi_y(S^1)$. But at the same time, as we shall see, the change of
basis simplifies the independence predicates $\IP(\C^*)$, making them
inversion-free, so that the computation of Equation~(\ref{eq:exp1})
can be reduced to evaluation of inversion-free queries. We now
formally define the change of basis. This consists of the
following steps: (1) we change the summation basis from $\bar T$ to $\bar
S$. (2) we translate the $\IP^n(\C^*)$ predicates from $\bar T$ to
$\bar S$. (3) we introduce a new set of predicates, called the link
predicates, on $\bar S$. (4) We add the remaining independence
predicates, $\IP^t(\C^*)$, translated from $\bar T$ to $\bar S$, to
$\bar S$.

Consider a factor $f \in \F^*$. It is a connected hierarchical query
with the hierarchy relation $\sqsubseteq$ on $Vars(f)$. Given $x \in
Vars(f)$, let $[x]$ denotes its equivalence class under $\sqsubseteq$
and let $\ceil{x}$ denote $\setof{y}{y \sqsupseteq x}$.
Define a {\em hierarchy tree} for $f$ as the tree where nodes are 
equivalence classes of variables, and edges are such that their
transitive closure is $\sqsubseteq$. For instance, in 
Example~\ref{ex:changebasis}, the hierarchy tree of $f$ has nodes
$\set{x},\set{y},\set{z}$ with $\set{x}$ as root and $\set{y},\set{z}$
its children.

Define a new vocabulary, consisting of a relation $S_f^{[x]}$ for each
$f \in \F^*$ and each node $[x]$ in the hierarchy tree of $f$, with
arity equal to the size of $\ceil{x}$. Let $\bar S$ denote
instances of this vocabulary. The intuition is that $S_f^{[x]}$ denotes
$\pi_{\ceil{x}}(T_f)$ in the change of basis from $\bar T$ to $\bar S$.
This completes step 1. 

Let $\IP^n$ denote the set of independence predicates on $\bar S$,
translated in a straightforward manner from the independence
predicates $\IP^n(\C^*)$ on $\bar T$ (details in appendix). This is
step 2.

Define a {\em link predicate} $S^{[x]}_f = \pi_{\ceil{x}}(S^{[y]}_f)$
for every edge $([x],[y])$ in the hierarchy tree of $f$. Let $\LP$ be
the set of all link predicates. This is step 3.

Finally, we add the trivial independence predicates $\IP^t$. For
this, we expand the basis of summation from $\bar S$ to $\bar S'$ by
adding the following sets. We add a new set $S^{i,j}_{x_i,x_j}$
corresponding to each pair $S_{f_i}^{[x_i]}$, $S_{f_j}^{[x_j]}$ such
that (i) $f_i$ and $f_j$ have a hierarchical join query which is
equivalent to $f_j$ and (ii) there are sub-goals $g_i$ in $h_i$ and
$g_j$ in $h_j$ referring to the same relation such that $Vars(g_i) =
\ceil{x_i}$ and $Vars(g_j) = \ceil{x_j}$. For each such
$S^{i,j}_{x_i,x_j}$, $\IP^t$ contains the following conjuncts:
$S_{f_i}^{[x_i]} \cap S_{f_j}^{[x_j]} = \emptyset$, $S^{i,j}_{x_i,x_j}
\subseteq S_{f_j}^{[x_j]}$. This describes the step 4.

Finally, we put it all together. We define a function $G(\bar S')$ on
$\bar S'$ as follows.  Consider a relation $S_f^{[x]}$, and let $p$ be
the number of children of $[x]$ in the hierarchy tree. For a tuple $t$
in $S_f^{[x]}$, let $$G(t) = (-1)^{p+1}\prod_{g \in sg(f) \mid Vars(g) =
\ceil{x}} p(g(t))$$ Define $G(\bar S') = \prod_{t \in \bar S'} G(t)$.

Denote $sig(\bar S')$ the set $\set{f \mid S_f^{[r_f]} \neq
\emptyset}$, where $[r_f]$ denotes the root of the hierarchy tree of
$f$.

\begin{theorem} With $\IP^t$, $\IP^n$, $\LP$, $sig$ and $G$ as defined above,
$$\sum_{\bar T \mid \IP(\C^*), sig(\bar T) = \sigma}  (-1)^{|\bar T|}
p(\F^*(\bar T)) = \sum_{\bar S' \mid \IP^n, \IP^t, \LP, sig(\bar S') =
\sigma} G(\bar S')$$
\end{theorem}

Finally, we use the bootstrapping principle to reduce the problem of
computing the summation to the evaluation of the query $\phi = (\IP^n
\wedge \IP^t \wedge \LP \wedge sig(\bar S') = \sigma)$.

\begin{lemma} The query $\phi$ defined above is an inversion-free
property.
\end{lemma}

By using Theorem~\ref{th:bootstraping}, we get the following:

\begin{theorem}
Suppose for every hierarchical join predicate $jp=\theta^T(h_i,h_j)$ between two factors
in $\H$, the join query $jq = \theta^R(f_i,f_j)$ 
has an eraser. Then, $q$ is \PTIME.
\label{th:dichotomy-ptime}
\end{theorem}

\section{\#P-Hard Queries}
\label{sec:inversion-hard}

Here we show the other half of Theorem~\ref{th:dichotomy}, i.e., if
$q$ has an inversion without an eraser, then $q$ is \#P-hard.

\eat{
We consider here a specific coverage for $q$. Fix a set of constants.
For each variable $x$ and a constant $a$, choose predicate $x = a$
or $x \neq a$. Also, for each pair of co-occurring variables $x$ and
$y$, choose a predicate $x = y$ or $x \neq y$. Let the covers consider
all possible queries that can be obtained this way, after remove the
unsatisfiable ones. Denote $\C = (\F,C)$ the resulting coverage. 
}

Let $\C = (\F,C, \bar x)$ be any strict coverage for $q$, 
$\C^*=(\F^*,C^*,\bar x^*)$ its closure and $\H$ the set of
hierarchical join queries over $\F$.

Suppose there are factors $h,h' \in \H$ such that the join
query $hj = \theta^T(h,h')$ has an inversion, but not an eraser. Among
all such $hj$, we will pick a specific one and use it to show that $q$
is \#\P-hard. Note that if there is no such $hj$, then the query is in
\PTIME\ by Theorem~\ref{th:dichotomy-ptime}.

Let the inversion in $hj$ consist of a unification path of length $k$
from $(f,x,y)$ with $x \sqsubset y$ to $(f',x',y')$ with $x' \sqsupset
y'$. Then, we will prove the \#\P-hardness of $q$ using a reduction
from the chain query $H_k$, which is \#\P-hard by Theorem~\ref{th:hard}.

Given an instance of $H_k$, we create an instance of $q$. The basic
idea is as follows: take the unification path in $hj$ that has the
inversion and completely unify it. We get a non-hierarchical query
(due to the inversion) with two distinguished variables $x$ and $y$
(the inversion variables), $k+2$ distinguished sub-goals (that
participated in the inversion), plus other sub-goals in the factor.
Use the structure of this query and the contents of the $k+2$
relations in the instance of $H_k$ to create an instance for $q$.
We skip the formal description of the reduction, but instead
illustrate it on examples.

\begin{example} 
\label{ex:hardness1}
Consider $q = U(x),V(x,y),V(y,x)$ and the coverage
$\C=(\F,C)$ where $\F = \set{f}$ with $f = U(x),V(x,y)$, $V(y,x), x \neq
y$ and $C = \set{\set{f}}$.  The coverage has a single factor and a
single cover. The first $V$ sub-goal of factor $f$ unifies with the
second sub-goal of another copy of $f$ to give an inversion between $x
\sqsupset y$ and their copy $y' \sqsubset x'$. If we 
unify the two sub-goals in two copies of $f$, we get the query:
$$q_u = {\underline U}(x),{\underline V}(x,y),V(y,x),{\underline U}(y)$$
We have underlined the sub-goals taking part in the inversion.
Now we give a reduction from the query $H_0 = R(x)$, $S(x,y),S(x',y'),T(y')$.
Given any instance of $R,S,T$ for $H_0$ construct an instance of $U,V$
as follows. We map the $R,S,T$ relations in $H_0$ to the
$U,V,U$ underlined sub goals of $q_u$ as follows: for each tuple
$R(a)$, create a tuple $U(a)$ with same probability.  For each
$S(a,b)$, create $V(a,b)$ with the same probability. For each $T(a)$,
create $U(a)$ with same probability. Also, for each $S(a,b)$,
create $V(b,a)$ with probability 1 (this corresponds to the
non-underlined sub-goal).

There is a natural 1-1 correspondence between the substructures of
$U,V$ and the substructures of $R,S,T$ with the same probability. It
can be shown that $q$ is true on a substructure iff the query
$R(x),S(x,y) \vee S(x',y'),T(y')$ is true on the corresponding
substructure. Thus, we can compute the probability of the query
$R(x),S(x,y) \vee S(x',y'),T(y')$, and hence, the probability of
$H_0$, by applying inclusion-exclusion.
\end{example}

Next, we show why a hardness reduction fails if the inversion has an
eraser. 

\begin{example} We revisit the query $q$ in
Example~\ref{ex:eraserquery2}. There
is an inversion between $x \sqsubset y$ in $f_1$ and $x' \sqsupset y'$
in $f_2$. However, their hierarchical join, $f_{12}$ have an eraser.
The unified query consists of $q_u=$\\
${\underline R}(r,x),{\underline S(r,x,y)},U(a,r),U(r,z),
V(r,z),V(a,r),{\underline T}(r,x)$\\
$R(a),S(a,b,c),U(a,a)$\\

We construct an instance $RSTUV$ for $q$ from an instance $R'S'T'$ for
$H_0$ as in previous example. However, there is a bad mapping from $q$
to $q_u$, corresponding to the eraser, which is $\set{r \rightarrow a,
x \rightarrow b, y \rightarrow c, x' \rightarrow x, y' \rightarrow y,
z \rightarrow r}$, which avoids the ${\underline R}$ sub-goal. The
effect is that $q$ is true on a world iff the query $S'(x',y')T'(y')$
(rather that $H_0$) is true on the corresponding world. So the
reduction from $H_0$ fails.
In fact, we know that this query $q$ is in \PTIME.
\end{example}

The final example shows that if there are multiple inversions without
erasers, we need to pick one carefully, which makes the hardness
reduction challenging.

\begin{example}
Consider the following variation of the query in previous example:
\begin{tabbing}
  $q = $\=$R(x),$\=$S(x,y),U(x,y,a,b),U(z_1,z_2,x,y),V(z_1,z_2,x,y)$ \\
        \>         \>$S(x',y'),T(y'),V(x',y',a,b)$ \\
        \>$R(a), S(a,b), U(a,b,a,b)$
\end{tabbing}
Let $f_1$ and $f_2$ denote the factors corresponding to the first two
lines of $q$. There is an inversion from $x \sqsupset y$ in $f_1$ to
$x' \sqsubset y'$ in $f_2$ via the two $S$ sub-goals, and it does not
have an eraser. But if we unify the two $S$ sub-goals to obtain
$S$, there is a "bad mapping" from $q$ to $q_u$ that maps $x,y$ to
$a,b$ and $z_1,z_2$ to $x,y$.  However, as it turns out, there is
another inversion in $q$ that we can use for hardness. The inversion is
from $x \sqsupset y$ to $z_1 \equiv z_2$ to $x',y'$ through the
following unification path: $U({\underline x}, {\underline y},x,y)$
unifies with (a copy of) $U({\underline z_1},{\underline z_2},x,y)$
and $V({\underline z_1},{\underline z_2},x,y)$ unifies with
$V({\underline x'},{\underline y'},a,b)$. We can show that this
inversion works for the hardness reduction.
\end{example}

By formalizing these ideas, we prove: 
\begin{theorem} Suppose there are $h, h' \in {\cal H}^*(q)$ such that 
their hierarchical join $hj$ has an inversion without an eraser. Then,
$q$ is $\#P$-complete.
\end{theorem}

\section{Conclusions}

We show that every conjunctive query has either \PTIME\ or
\#\P-complete complexity on a probabilistic structure. As part of the
analysis required to establish this result we have introduced new
notions such as hierarchical queries, inversions, and erasers.  Future
work may include several research directions: a study whether the
hardness results can be sharpened to counting the number of
substructures (i.e.  when all probabilities are 1/2); an analysis of
the query complexity; extensions to richer probabilistic models (e.g.
to probabilistic databases with disjoint and independent
tuples~\cite{dalvi06debul}); and, finally, studies for making our
\PTIME\ algorithm practical for probabilistic database systems.

{

\small
\bibliographystyle{plain}
\bibliography{/projects/xmltk/nileshIndexing/master.bib}
}

\onecolumn
\appendix

\eat{
{\bf DAN, IS THIS OBSOLETE NOW?}

We first show $\succeq$ is
transitive.  Consider a path from $(qc,x,y)$ to $(qc',x',y')$ with $x'
\sqsupset y'$ and a path $P$ from $(qc,y,z)$ with to $(qc'',y'',z'')$
$y'' \sqsupset z''$.  $P$ starts with a subgoal $g$ in $qc$ containing
$y,z$: that subgoal also contains $x$, since $x \sqsupseteq y$, hence
the next subgoal on $P$ has a variable that unifies with $x$.
Continuing this argument, every subgoal used on $P$ has variable that
connects to $x$, hence we conclude that $qc'''$ has a variable $x'''$
s.t. $x'' \sqsupseteq y'' \sqsupset z''$.  This gives us a path
$(qc,x,z)$ to $(qc'',x'',z'')$ hence $x \succeq z$.
The extension condition follows immediately from the fact
that $q$ is inversion free.  Now we verify the three conditions.
Extension: if $x \succeq y$ then $x,y$ co-occur, and $x \sqsubset y$
is not possible because it would lead to an inversion, hence $x
\sqsupset y$.  Total: follows immediately by taking the empty path
from $(x,y)$ to $(x,y)$.  Monotonicity: follows from the fact that if
$x \succeq y$ unifies with $x' \prec y'$ then we can construct an
inversion.
}

\section{Examples of Inversions}
\label{app:examples}

We illustrate in Fig.~\ref{fig:inversion-free} several subtleties of
inversion-free queries that were left out from the text.
Fig.~\ref{fig:inversion} illustrates some queries with inversions; all
are \#\P-hard.

\begin{figure*}[h]
\small
  \centering
  \begin{tabular}{|l|l|l|} \hline\hline
Query.  The trivial coverage & Fragment of  a strict coverage & Comments \\
is non-strict and has an ``inversion'' & (Unification chain underlined) & \\
\hline\hline
\parbox{4cm}{
  \begin{eqnarray*}
    &   & R(x)S_1(\underline{x},y,\underline{y}) \\
    &   & S_1(\underline{u},v,\underline{w}),S_2(\underline{u},v,\underline{w}) \\
    &   & S_2(\underline{x'},x',\underline{y'}),T(y')
  \end{eqnarray*}
}
&
\parbox{6cm}{
  \begin{eqnarray*}
    qc_1 & = & R(x),S_1(\underline{x},y,\underline{y}), x\not=y, \\
         &   & S_1(\underline{u},v,\underline{v}),S_2(\underline{u},v,\underline{v}),u\not=v \\
         &   & S_2(x',x',y'),T(y'),x'\not=y' \\
    qc_2 & = & R(x),S_1(x,y,y), x\not=y\\
         &   & S_1(\underline{u},u,\underline{w}),S_2(\underline{u},u,\underline{w}),u\not=w \\
         &   & S_2(\underline{x'},x',\underline{y'}),T(y'),x'\not=y'
  \end{eqnarray*}
} &
\parbox{5cm}{Illustrates the need for a strict coverage.  The
  unification path forming an inversion in  $q$ in the trivial cover
  (which is non-strict) is interrupted when
  we add  $\not=$ predicates to make the cover strict.
}
\\ \hline
\parbox{4cm}{
  \begin{eqnarray*}
     &   & R(x_1,x_2), S(\underline{x_1},x_2,\underline{y},y), \\
     &   & S(x_1,x_1,x_2,x_2) \\
     &   & S(\underline{x'},x',\underline{y'},y'),T(y') \\
  \end{eqnarray*}
}
&
\parbox{6cm}{
  \begin{eqnarray*}
    qc & = & R(x,x), S(\underline{x},x,\underline{y},y), \\
       &   & S(x,x,x,x), x\not= y \\
       &   & S(\underline{x'},x',\underline{y'},y'), T(y'),x'\not=y' \\
       & = & R(x,x),S(x,x,x,x), \\
       &   & S(x',x',y',y'), T(y'),x'\not=y'
  \end{eqnarray*}
}
&
\parbox{5cm}{This illustrates the need to minimize covers.  The
  inversion disappears after minimizing  $qc$.
}
\\ \hline
\parbox{4cm}{
  \begin{eqnarray*}
      &   & R(x_1,x_2),S(\underline{x_1},x_2,\underline{y},y) \\
      &   & S(x_1,x_2,x_1,x_2) \\
      &   & S(\underline{x'},x',\underline{y_1'},y_2'),T(y_1',y_2')
  \end{eqnarray*}
}
&
\parbox{6cm}{
  \begin{eqnarray*}
    qc_1 & = & R(x,x),S(\underline{x},x,\underline{y},y), x\not=y\\
         &   & S(\underline{x'},x',\underline{y'},y'),T(y',y'),x'\not=y' \\
         &   & S(x,x,x,x) \\
    qc_2 & = & R(x,x), S(x,x,x,x), \\
         &   & S(x',x',y',y'),T(y',y'),x'\not=y'
  \end{eqnarray*}
}
&
\parbox{5cm}{This shows that we should not  consider redundant
  coverages.   There is an inversion in $qc_1$, but this cover is
  contained in $qc_2$ so it is redundant and after we remove $qc_1$
  from the coverage there is no more inversion.
} 
\\ \hline\hline
  \end{tabular}
\caption{Inversion-free queries: all are in \PTIME. \eat{ This illustrates
  several subtle points in the definition of an inversion, which are
  critical in order to make the \PTIME\ algorithm applicable to these
  queries.}}
  \label{fig:inversion-free}
\end{figure*}

\begin{figure*}[h]
\small
  \centering
  \begin{tabular}{|l|l|l|} \hline\hline
Query & Fragment of a strict coverage  (inversion underlined) &
Comments \\ \hline\hline
\parbox{3.5cm}{
  \begin{eqnarray*}
    & & R(x,y),R(y,z)
  \end{eqnarray*}
} &
\parbox{7cm}{
  \begin{eqnarray*}
    qc  & = & R(x,y),R(\underline{y},\underline{z}) \\
    qc  & = & R(\underline{x'},\underline{y'}),R(y',z')
  \end{eqnarray*}
} &
\parbox{5cm}{Here and the inversion is between $y \sqsupset z$
  and $x' \sqsubset y'$ in a copy of itself.}
\\ \hline
\parbox{3.5cm}{
  \begin{eqnarray*}
      &   & R(x),S_1(x,y), \\
      &   & S_1(u_1,v_1),S_2(u_1,v_1) \\
      &   & S_2(u_2,v_2),S_2(v_2,u_2)
  \end{eqnarray*}
} &
\parbox{7cm}{
  \begin{eqnarray*}
    qc_1 & = & R(x),S_1(\underline{x},\underline{y}), x>y, \\
      &   &
      S_1(\underline{u_1},\underline{v_1}),S_2(\underline{u_1},\underline{v_1}),
      u_1>v_1 \\
      &   & S_2(\underline{u_2},\underline{v_2}),S_2(v_2,u_2),  u_2>v_2 \\
    qc_2 & = & R(x),S_1(\underline{x},\underline{y}), x<y, \\
      &   &
      S_1(\underline{u_1},\underline{v_1}),S_2(\underline{u_1},\underline{v_1}),
      u_1<v_1 \\
      &   & S_2(u_2,v_2),S_2(v_2,u_2),  u_2<v_2
  \end{eqnarray*}
} &
\parbox{5cm}{Here $x \sqsupset y$, $u_1 \equiv v_1$, $u_2 \equiv v_2$
  and the inversion path goes twice through each factor.  We call this
an {\em open marked ring}.}
\\ \hline
\parbox{3.5cm}{
  \begin{eqnarray*}
    & & R(x),S(x,y),S(y,x)
  \end{eqnarray*}
} &
\parbox{7cm}{
  \begin{eqnarray*}
    qc_1 & = & R(x),S(\underline{x},\underline{y}),S(y,x),x<y \\
    qc_2 & = & R(x'),S(x',y'),S(\underline{y'},\underline{x'}),x'>y' \\
  \end{eqnarray*}
} &
\parbox{5cm}{Here $x \sqsupset y$ and the inversion is between $x,y$
  and their copy $y',x'$.  We call this a {\em marked ring}.}
\\ \hline
\parbox{3.5cm}{
  \begin{eqnarray*}
      &   & R(x),S_1(x,y), \\
      &   & S_1(u_1,v_1),S_2(u_1,v_1) \\
      &   & S_2(u_2,v_2),S_2(v_2,u_2)
  \end{eqnarray*}
} &
\eat{
\parbox{7cm}{
  \begin{eqnarray*}
    qc_1 & = & R(x),S_1(\underline{x},\underline{y}), x>y, \\
      &   &
      S_1(\underline{u_1},\underline{v_1}),S_2(\underline{u_1},\underline{v_1}),
      u_1>v_1 \\
      &   & S_2(\underline{u_2},\underline{v_2}),S_2(v_2,u_2),  u_2>v_2 \\
    qc_2 & = & R(x),S_1(\underline{x},\underline{y}), x<y, \\
      &   &
      S_1(\underline{u_1},\underline{v_1}),S_2(\underline{u_1},\underline{v_1}),
      u_1<v_1 \\
      &   & S_2(u_2,v_2),S_2(v_2,u_2),  u_2<v_2
  \end{eqnarray*}
} &
\parbox{5cm}{Here $x \sqsupset y$, $u_1 \equiv v_1$, $u_2 \equiv v_2$
  and the inversion path goes twice through each factor.  We call this
an {\em open marked ring}.}
\\ \hline
\parbox{3.5cm}{
  \begin{eqnarray*}
     &   & R(x), S(x,y,y), \\
     &   & T(u,v), S(u,v,w), \\
     &   & U(y'),S(x',y',x')
  \end{eqnarray*}
}
&
}
\parbox{7cm}{
  \begin{eqnarray*}
   qc_1 & = & R(x), S(\underline{x},\underline{y},y), x \not= y, \\
          & & T(\underline{u},\underline{v}),
          S(\underline{u},\underline{v},v), u \not= v,\\
          & & U(y'),S(x',y',x'), x' \not= y' \\
   qc_2 & = & R(x), S(x,y,y), x \not= y\\
        &   & T(\underline{w},\underline{v}),
          S(\underline{w},\underline{v},w), w \not=v, \\
        &   & U(y'),S(\underline{x'},\underline{y'},x'), x'\not=y'
  \end{eqnarray*}
} &
\parbox{5cm}{Here the inversion path goes twice through the subgoal
  $S(u,v,w)$ using different pairs of variables.}
\\ \hline\hline
  \end{tabular}
 \caption{Queries with inversions: all are \#P-hard}
  \label{fig:inversion}
\end{figure*}

\section{Proof of Theorem~1.4}
\label{app:nonhierarchical}

Let $P$ be a conjunctive formula and \struct{A} be a structure.  We
say that $P$ is {\em decisive} w.r.t. \struct{A} if there exists a
function $c : A \rightarrow Var(P)$ s.t. for any homomorphism $h : P
\rightarrow \struct{A}$ there exists an automorphism $i : P
\rightarrow P$ s.t. denoting $h' = h \circ i$ we have $c \circ h' =
id_P$.  The function  $c$, which we call a {\em choice} function,
``chooses'' for each node $u$ in $A$ a variable $x = c(u)$ in $P$ such
that any homomorphism from $P$ to $\struct{A}$ maps $x$ to $u$, up to
renaming of variables in $P$.  Let $S$ be a class of structures.  We
say that $P$ is decisive w.r.t. $S$ if it is decisive w.r.t. to each
structure in $S$.

In the sequel we will make use of the following two classes of graphs.
A {\em 4-partite graph} has nodes partitioned into four classes $V_i$,
$i=1,2,3,4$, and edges are subsets of $\bigcup_{i=1}^3 V_i \times
V_{i+1}$.  A {\em triangled-graph} has a distinguished node $v_0$ and
two disjoint sets of nodes $V_1, V_2$ s.t. edges are subsets of
$(\set{v_0} \times V_1) \cup (V_1 \times V_2) \cup (V_2 \times
\set{v_0})$.

\begin{example} \label{ex:decisive1}
The query below checks if a graph has a chain of length 3:
\begin{verbatim}
 P_3 =  E(x,y), E(y,z), E(z,u)
\end{verbatim}
Then $P_3$ is decisive on the set of 4-partite graphs.  To see this,
the choice function simply chooses to map $V_1$ to $x$, $V_2$ to $y$,
$V_3$ to $z$ and $V_4$ to $u$.

\end{example}

\begin{example}  \label{ex:decisive2}

The query below checks if the graph has a triangle:
\begin{verbatim}
 T = E(x,y), E(y,z), E(z,x)
\end{verbatim}
Then $T$ is decisive on the class of triangled graphs.  To see this,
consider a triangled graph $G$ and define $c$ to map $v_0$ to $x$,
$V_1$ to $y$ and $V_2$ to $z$.  A homomorphism $h : T \rightarrow G$
may map $x$ to some other node than $v_0$, but after a proper rotation
(automorphism) we transform $h$ into a homomorphism $h \circ i$ that
is consistent with $c$.

Note that $T$ is not decisive on the class of all graphs.  For example
it is not decisive on the complete graph $K_4$.  

\end{example}

Our interest in the two queries above and their associated classes of
decisive structures comes from the fact that their complexity is
\#\P-complete:

\begin{proposition}
  Let $P_3$ be the 3-chain property in Example~\ref{ex:decisive1}.
  The complexity of computing $\Pr[P_3]$ on 4-partite graphs is
  \#\P-complete.
  
  Let $T$ be the triangle property in Example~\ref{ex:decisive2}.  The
  complexity of computing $\Pr[T]$ on triangled graphs is
  \#\P-complete.
\end{proposition}

\begin{proof}
  By reduction from the problem of computing the probability of
  bipartite 2DNF formulas.  Let $X = \set{x_1, \ldots, x_m}$ and $Y =
  \set{y_1, \ldots, y_n}$ be two disjoint sets of Boolean variables,
  and consider a bipartite 2DNF formula:
  \begin{eqnarray}
    \Phi & = & \bigvee_{k=1,t} x_{i_k} \wedge y_{j_k} \label{eq:phi}
  \end{eqnarray}
  Construct the following 4-partite graph: $V_0 = \set{u}$, $V_1 = X$,
  $V_2 = Y$, $V_4 = \set{v}$, where $u, v$ are two new nodes.  All
  edges from $u$ to $x_i$ are present and their probability if
  $\Pr[x_i]$; for each clause $x_{i_k} \wedge y_{j_k}$ in
  (\ref{eq:phi}) there is an edge $(x_{i_k}, y_{j_k})$ with
  probability 1, and all edges $(y_j, v)$ are present and have
  probability $\Pr[y_j]$.  Clearly the probability that this graph has
  a path of length 3 is precisely $\Pr[\phi]$.  This proves the
  hardness of $P_3$.  The hardness of $T$ is obtained similarly, by
  merging $u$ and $v$ into a single node.
\end{proof}

\begin{theorem}
  Let $Q$ be a conjunctive formula, which is minimal, and let $P$ be
  subformula.  If there exists a class of structures $S$ s.t. (1) $P$
  is decisive on $S$ and (2) $P$ is \#\P-complete on $S$, then $Q$
  is \#\P-complete on the class of all structures.
\end{theorem}

\begin{proof}
  We reduce the problem of evaluating $P$ on some structure in $S$ to
  the problem of evaluating $Q$ on an arbitrary structure.  Let
  $\struct{A} \in S$, and $c : A \rightarrow Var(P)$ be a choice
  function.  We construct a new structure $\struct{B}$ as follows.
  First define $H = \setof{h : P \rightarrow \struct{A}}{c \circ h =
    id_P}$ to be the set of homomorphism from $P$ to $\struct{A}$ that
  are consistent with the choice function.  Note that $H$ is
  polynomial in the size of $\struct{A}$ since $P$ is fixed.  Define
  the new structure $\struct{B}$ as follows.  Its nodes, $B$ are
  obtained as follows.  First define the set $N = \setof{(x,h)}{x \in
    Var(Q), h \in H}$; next define the equivalence relation $(x,h)
  \equiv (x',h')$ if $(x,h) = (x',h')$, or if $x=x' \in Var(P)$ and
  $h(x) = h'(x)$ (i.e.  collapse multiple copies of the same variable
  from $P$ if they are mapped to the same node in $A$).  The nodes in
  $\struct{B}$ are equivalence classes $[(x,h)]$, i.e. $B = N
  /\equiv$.  The relations in $\struct{B}$ are of the form
  $R([(x_1,h)], \ldots, [(x_k,h)])$, where $R(x_1, \ldots, x_k)$
  appears in $Q$, and $h \in H$.  One can think of $\struct{B}$ as
  consisting of multiple copies of $Q$, one for each possible way of
  mapping $P$ into $\struct{A}$, but such that all copies of the same
  $P$-variable that are mapped to the same node $u \in A$ are merged
  into a single node. The latter are precisely the nodes of the form
  $[(x,h)]$ for $x \in Var(P)$, and we call them the {\em special
    nodes} in $\struct{B}$. Thus, the special nodes in $\struct{B}$
  form a substructure that is isomorphic to some substructure
  $\struct{A}_0$ of $\struct{A}$, which is large enough to contain the
  image of all homomorphism from $P$ to $\struct{A}$.  The
  probabilities are as follows.  If $x_1, \ldots, x_k \in Var(P)$ then
  $$\Pr_B(R([(x_1,h)], \ldots, [(x_k,h)])) = \Pr_A(R(h(x_1), \ldots,
  h(x_k)))$$; otherwise $\Pr_B(R([(x_1,h)], \ldots, [(x_k,h)])) = 1$.
  Note that there is a 1-to-1 correspondence between the worlds $W_A$
  of $\struct{A}$ and the worlds $W_B$ of $\struct{B}$, and $\Pr[W_A]
  = \Pr[W_B]$.

  Claim 1.  Let $W_A$ be a world of $\struct{A}$ s.t. $W_A \models P$.
  Then, denoting $W_B$ the corresponding world of $\struct{B}$, we
  have $W_B \models Q$.  Indeed, let $h : P \rightarrow \struct{A}$ be
  a homomorphism whose image uses only tuples in $W_A$.  We can assume
  w.l.o.g. that it is consistent with the choice function, i.e.  $c
  \circ h = id_P$ (otherwise simply compose it with the automorphism
  $i$), hence $h \in H$.  Extended it to a homomorphism $\bar h : Q
  \rightarrow \struct{B}$ by defining $\bar h(x) = [(x,h)]$: it
  clearly only uses tuples in $W_B$.
  
  Claim 2. Let $W_B$ be a world of $\struct{B}$ s.t. $W_B \models Q$.
  Then, denoting $W_A$ the corresponding world of $\struct{A}$ we have
  $W_A \models P$.  Let $\bar h : Q \rightarrow \struct{B}$ be a
  homomorphism.  If $\bar h$ maps $Var(P)$ only to the special nodes
  in $\struct{B}$, then we are done; but this may not necessarily be
  the case.  We will prove instead that there exists some automorphism
  $g : Q \rightarrow Q$ s.t. $\bar h \circ g$ maps $Var(P)$ to the
  special nodes in $\struct{B}$.
  
  Define the function $f : B \rightarrow Var(Q)$ to be $f([x,h]) = x$;
  one can check that it is a homomorphism from $\struct{B}$ to $Q$,
  and that all special nodes and only these are mapped to $Var(P)$.
  Consider the composition $f \circ \bar h : Q \rightarrow Q$, which
  is an isomorphism (since $Q$ is minimal); in particular $\bar
  h^{-1}$ is functional, i.e. $|\bar h^{-1}(u)| \leq 1$.  Define $g =
  (f \circ \bar h)^{-1}$ to be its inverse.  Then $\bar h \circ g$
  maps $Var(P)$ to the special nodes in $\struct{B}$.  Indeed, for any
  variable $x \in Var(P)$, $f^{-1}(x)$ consists only of special nodes,
  hence $\bar h(g(x)) = \bar h(\bar h^{-1}(f^{-1}(x))) = Dom(h^{-1})
  \cap f^{-1}(x)$ is a special node.
\end{proof}

\begin{theorem}
  Let $P = R_1(\bar v_1), R_2(\bar v_2), R_3(\bar v_3)$ be a
  conjunctive property, which is minimal, and for which there exists
  two variables $x, y$ s.t. $x \in \bar v_1, x \in \bar v_2, x \not\in
  \bar v_3$ and $y \not\in \bar v_1, y \in \bar v_2, y \in \bar v_3$.
  Then there exists a class of structures $S$ s.t. (a) $P$ is decisive
  w.r.t. $S$ and (b) $P$ is \#\P-complete on structures in $S$.
  Note that $R_1, R_2, R_3$ may be any relation names, possibly the
  same relation name.
\end{theorem}

\begin{proof}
  By reduction from partitioned 2DNF.  Consider Eq.(\ref{eq:phi}), and
  recall that the variables are $X = \set{x_1, \ldots, x_m}$,
  $Y = \set{y_1, \ldots, y_n}$.  Let $U = \set{u_1, u_2, \ldots, u_k}$
  be all the variables occurring in $P$ in addition to $x$ and $y$,
  and $C$ be the set of constants.  Define the structure $\struct{A}$
  s.t. $A = X \cup Y \cup U \cup C$, and the relations are defined as
  follows:

  \begin{eqnarray*}
    R^A_1 & = & \setof{R_1(\bar v_1[x_i / x])}{i=1,m} \\
    R^A_2 & = & \setof{R_1(\bar v_1[x_{i_k} / x, y_{j_k}] / y)}{k=1,t} \\
    R^A_3 & = & \setof{R_3(\bar v_3[y_j / y])}{j=1,n}
  \end{eqnarray*}
  Thus, the tuples in the first set correspond to the Boolean
  variables $x_i$, those in the second set correspond to clauses
  $x_{i_k} \wedge y_{j_k}$, and those in the third set correspond to
  the Boolean variables $y_j$.  Note that the three sets defined on
  the right are disjoint: if two or more of the relation names $R_1,
  R_2, R_3$ are the same, then their interpretation in $\struct{A}$
  consists of the union of the corresponding right hand definitions
  above.  The tuple probabilities are as follows: those in $R^A_1$ are
  precisely $\Pr(x_i)$, those in $R^A_2$ are 1, and those in $R^A_3$
  are precisely $\Pr(y_j)$.
  
  We first show that $P$ is decisive on $\struct{A}$.  Define the
  choice function $c : A \rightarrow Var(P)$ to be $c(x_i) = x$ for
  $i=1,m$, $c(y_j) = y$ for $j=1,n$ and $c(u_p) = u_p$ for $p=1,k$.
  We need to prove that every homomorphism $h : P \rightarrow
  \struct{A}$ is, up to isomorphism, consistent with the choice
  function.  For that we note that the choice function itself is a
  homomorphism $c : \struct{A} \rightarrow P$, hence $c \circ h : P
  \rightarrow P$ is an automorphism (since $P$ is minimal), and we
  denote $i = (c \circ h)^{-1}$.  We show now that $h' = h \circ i$ is
  consistent with $c$.  Indeed: $c \circ h' = c \circ h \circ (c \circ
  h)^{-1} = id_P$.

Next we prove that the probability of $P$ being true on $\struct{A}$
is the same as the probability that $\Phi$ is true.  There is an
obvious one-to-one correspondence between worlds $W_A$ of $\struct{A}$
and truth assignment for $\Phi$: the tuple in $R^A_1$ corresponding to
$x_i$ occurs in $W_A$ iff $x_i = true$, and similarly for $R^A_3$ and
the $y_j$'s.  Clearly if the truth assignment makes $\Phi$ true, then
$P$ is true on $W_A$: simply pick two variables $x_i$ and $y_j$ that
are both true under the truth assignment, and note that $P$ can be
mapped to the three tuples corresponding to $x_i$, to the clause $x_i
\wedge y_j$ and to $y_j$ respectively.  Conversely, suppose $P$ is
true on $W_A$, i.e. there exists a homomorphism $h : P \rightarrow
\struct{A}$ whose image is contained in $W_A$.  Since $P$ is decisive
on $\struct{A}$ there exists another homomorphism $h' : P \rightarrow
\struct{A}$ that is consistent with $c$, i.e. it maps $x$ to some
$x_i$ and $y$ to some $y_j$.  Then $Im(h')$ consists of three tuples
$R_1^A(\bar v_1[x_i / x])$, $R_2(\bar v_2[x_{i_k} / x, y_{j_k} / y])$,
and $R_3^A(\bar v_3[y_j / y])$, and, moreover $x_{i_k} \wedge y_{j_k}$
is a clause in $\Phi$, which is true under the truth assignment
corresponding to $W_A$.

\end{proof}

\eat{
\begin{corollary}
  Let $Q$ be a conjunctive property, which is minimal, and which
  contains a subformula $P = R_1(\bar v_1), R_2(\bar v_2), R_3(\bar
  v_3)$ which is also minimal, and for which there exists two
  variables $x$, $y$ s.t.  $x \in \bar v_1, x \in \bar v_2, x \not\in
  \bar v_3$ and $y \not\in \bar v_1, y \in \bar v_2, y \in \bar v_3$.
  Then $Q$ is \#\P-complete.
\end{corollary}
}
\begin{corollary}
  Let $Q$ be a non-hierarchical conjunctive query. Then $Q$ is
  \#\P-hard.
\end{corollary}

\begin{proof} Consider the minimal conjunctive query defined by $Q$.
Since $Q$ is non-hierarchical, there must be two variables $x$ and $y$
such that $sg(x) \cap sg(y) \neq \emptyset$, $sg(x) - sg(y) \neq
\emptyset$ and $sg(y) - sg(x) \neq \emptyset$. Thus, the minimal query
must contain a subformula $P = R_1(\bar v_1), R_2(\bar v_2), R_3(\bar
v_3)$ s.t.  $x \in \bar v_1, x \in \bar v_2, x \not\in \bar v_3$ and
$y \not\in \bar v_1, y \in \bar v_2, y \in \bar v_3$.

It follows from the previous two results that $Q$ is \#\P-hard.
\end{proof}

\section{Proof of Theorem~1.5}

We will prove here that for every $k \geq 0$, $H_k$ is \#\P-hard.
Recall that 
  \begin{tabbing}
    $H_k$ =\\
    $      R(x),$\=$S_0(x,y),$ \\
     \>$S_0(u_1,v_1),$\=$S_1(u_1,v_1)$ \\
     \>       \>$S_1(u_2,v_2)$,\ldots\= \\
     \>       \>      \>$S_{k-1}(u_k,v_k),$\=$S_k(u_k,v_k)$ \\
     \>       \>      \>     \>$S_k(x',y'),T(y')$
  \end{tabbing}

Define queries $\phi_0, \cdots, \phi_{k+1}$, where 
\begin{eqnarray*}
\phi_0 &=& R(x),S_0(x,y)\\
\phi_i &=& S_{i-1}(u,v),S_i(u,v)~\mbox{for $1 \leq i \leq k$}\\
\phi_{k+1} &=& S_k(x',y'),T(y')
\end{eqnarray*}
Thus,
$H_k = \bigwedge_{i \in [k]} \phi_i$. For any proper subset $S$ of
$[k]$, the query $\bigwedge_{i \in S}$ is in \PTIME (this follows from
a result we prove later that every inversion-free query is in \PTIME). 
Using the principle of inclusion-exclusion, to show the hardness of
$H_k$, it is enough to show the hardness of the query $\bigvee_{i
\in [k]} \phi_i$ is hard, or equivalently, its negation $q =
\bigwedge_{i \in [k]} (NOT \phi_i)$.

We give a reduction from the problem of computing the probability of
bipartite 2DNF formulas.  Let $X = \set{x_1, \ldots, x_m}$ and $Y =
\set{y_1, \ldots, y_n}$ be two disjoint sets of Boolean variables, and
consider a bipartite 2DNF formula:
  \begin{eqnarray}
    \Phi & = & \bigvee_{h=1,t} x_{i_h} \wedge y_{j_h}
  \end{eqnarray}

We construct an instance for relations $R, S_0, \cdots, S_k, T$.
For each variable $x_i \in X$, create a tuple $R(x_i)$ and assign it a 
probability 1/2. For each $y_i \in Y$, create a tuple $T(y_i)$ and
assign it a probability 1/2.  For each clause $(x_{i_h}, y_{j_h})$,
and for each $l \in [k]$, create a tuple $S_l( (x_{i_h}, y_{j_h})$ and
assign it a probability $p_1$ for $l = 0,k$ and a probability of $p_2$
for $1 \leq l \leq k-1$.
 
Let $T_{i,j}$ be the number of assignments of $\Phi$ such that $i$
clauses have both variables true and $j$ clauses have no variables
true. Thus, $(t - i - j)$ have exactly 1 variable true, where $t$ is
the number of clauses.

There is a canonical mapping between the truth assignments of $X,Y$
and worlds of relations $S,T$ where $x \in X$ is true iff $S(x)$ is
present and $y \in Y$ is true iff $T(y)$ is present.

Consider some fixed assignment where $i$ clauses have both variables
true and $j$ clauses have no variables true. Fix relations $R,T$
accordingly and consider all possible worlds of $S_1, \cdots, S_k$
such that $q$ is true on the worlds. For each $(x_{i_h}, y_{j_h})$,
consider all tuples of the form $S_l(x_{i_h}, y_{j_h})$:
\begin{enumerate}
\item If $x_{i_h}$ and $y_{i_h}$ are true, the tuples $S_0(x_{i_h},
y_{j_h})$ must be both out, and other edges do not matter. Its
probability is $(1-p_1)^2$
\item If one of them is true, one of the tuples $S_0(x_{i_h}, 
y_{j_h})$ must be out (depending on which variable is true), and other
edges do not matter. Its probability is $(1-p_1)$.
\item If $x_{i_h}$ and $y_{i_h}$ are both false, the only requirement
is that not all $S_l(x_{i_h}, y_{j_h})$ are in. Its probability is $(1
- p_1^2p_2^{k-2})$.
\end{enumerate}

Thus, its total probability of all worlds corresponding to this fixed
assignment is
$$(1/2)^{|X|+|Y|} [(1-p_1)^2]^i  [(1 - p_1^2p_2^{k-2})]^j [(1-p_1)]^{c -
i - j}$$
This can be written as $KA^iB^j$, where $K = (1/2)^{|X|+|Y|}(1 -
p)^c$, $A=(1 - p_1)$ and $B = (1 -  p_1^2p_2^{k-2})/(1-p_1)$.

Thus $Pr[q] = \sum_{i,j : i + j \leq t}  T_{i,j} K. A^i . B^j$

This is a linear equation in variables $T_{i,j}$. We put different
values of $p_1,p_2$ to get different values of $A$,$B$ and get a
system of linear equations. The coefficient matrix of this set of
equations is the Vandemonte matrix which is known to be
invertible. By inverting the matrix, we solve for
each $T_{i,j}$. Finally, we can compute the number of satisfying
assignments of $\phi$ using $\sum_{i,j \mid i+j \leq t, j \neq t}
T_{i,j}$. This gives a polynomial time reduction from the problem of
computing $H_k$ to counting the number of satisfying assignments of a
bipartite DNF formula. Hence, $H_k$ is \#\P-hard.

\section{Proof of Theorem~2.7}

\label{sec:pie}
Consider some probability space. Let $\vec{U} = (U_1, \cdots, U_k)$ be
a vector consisting of $k$ sets. For each $i \in [k]$ and each $x \in
U_i$, let $E(i,x)$ be an event in the probability space. Define $E(i)
= \bigvee_{x \in U_i} E(i,x)$. Let $Q$ be a {\tt CNF} formula over
events $E(1), \cdots E(k)$, i.e., let $\psi$ be a set of subsets of
$[k]$ and let 
\begin{equation}
Q = \bigvee_{S \in \psi} \bigwedge_{i \in S} E(i)
\label{eq:qdef}
\end{equation}
We will derive an expression for $Pr[Q]$ in terms of the probabilities
of the events $E(i,x)$. We need some notations. A {\em signature} is
simply a subset of $[k]$.  Given a vector of sets $\vec{S} = (S_1,
\cdots, S_k)$, the signature of $\vec{S}$, denoted $sig(\vec{S})$, is
the set $\{i \mid S_i \neq \emptyset\}$. $E(\vec{S})$ is defined as
the event $\bigwedge_{i \in [k]} E(i, S_i)$. The size of $\vec{U}$
is defined as $|\vec{U}| = |U_1| + \cdots + |U_k|$. Also, given
vectors $\vec{S}$ and $\vec{T}$, we say that $\vec{S} \subseteq
\vec{T}$ iff for all $i \in [k]$, $S_i \subseteq T_i$. 

Define the upward closure of $\psi$ as $\UP(\psi) = \{sg \mid sg
\subseteq [k], \exists sg_0 \in \psi~s.t.~ sg_0 \subseteq sg\}$.
Define the minimal elements of $\psi$ as $\factors(\psi) = \{sg \mid
sg \in \psi, \forall sg_0 \in \psi.~sg_0 \subseteq sg \Rightarrow sg_0
= sg\}$. For a set of signatures $G$, let $sig(G) = \cup_{sg \in G}
sig(sg)$.  Given a signature $sg$, define 
$$N(sg) = (-1)^{|sg|} \sum_{G \mid G \subseteq \factors(\psi), sig(G)
= sg} (-1)^{|G|}$$
Our main result is follows:

\begin{theorem} With $\vec{U}$, $\psi$ and $Q$ as defined above,
$$Pr[Q] = \sum_{\vec{S} \subseteq \vec{U}} N(sig(\vec{S})) (-1)^{|\vec{S}|}$$
\label{th:pie}
\end{theorem}

We will need the following result later which gives an alternate
formula for $N(sg)$.

\begin{lemma} $N(sg) = \sum_{\{sg_0 \mid sg_0 \subseteq sg, sg_0 \not\in
UP(\psi)\}} (-1)^{|sg_0|}$.
\end{lemma}

In the rest of the section, we prove this theorem.

Let $*$ be an element such that $*
\not\in U_i$ for all $i$ and define $U_i^* = U_i \cup \{*\}$. Given an
element $x \in U_1^* \times \cdots \times U_k^*$, the {\em signature} of
$x$, denoted $sig(x)$, is a subset of $[k]$ given by $\{i \mid
\pi_i(x) \neq *\}$. Given a vector of sets $\vec{S} = (S_1, \cdots,
S_k)$ where $S_i \subseteq U_i$, define 
$$\Pi_{\psi}(\vec{S}) = \{ x \mid x \in (S_1 \cup \{*\}) \times \cdots
(S_k \cup \{*\}), sig(x) \in \psi\}$$
Given a vector $\vec{x} \in \Pi_{\psi}(\vec{U})$, define $E(\vec{x})
= \bigwedge_{i \in sig(x)} E(i, \pi_i(x))$.  
Then, from Eq~(\ref{eq:qdef}), it follows that
$$
Q = \bigvee_{\vec{x} \in \Pi_{\psi}(\vec{U})} E(\vec{x})
$$
Using inclusion-exclusion, we obtain
\begin{equation}
Pr[Q] = \sum_{T \subseteq \Pi_{\psi}(\vec{U})} (-1)^T Pr[ \bigwedge_{x \in T} E(x)]
\label{eq:qexpand}
\end{equation}

For a set $T \subseteq \Pi_{\psi}(\vec{U})$, define $\pi_i(T) =
\{\pi_i(x) \mid x \in T, \pi_i(x) \neq *\}$. Also, define $E(i, S) =
\bigwedge_{s \in S} E(i, s)$. Then, $\bigwedge_{x \in T} E(x) = E(1,
\pi_1(T)) \wedge \cdots \wedge E(k, \pi_k(T))$.

In Eq~(\ref{eq:qexpand}, we group the $T$ based on their projection to
obtain
\begin{equation}
Pr[Q] = \sum_{S_1, \cdots, S_k} Pr[ \bigwedge_{i \in [k]} E(i, S_i)] *
(\sum_{T \subseteq \Pi_{\psi}(\vec{U}), \pi_i(T) = S_i} (-1)^T)
\label{eq:qgroup}
\end{equation}
Let $N(S_1, \cdots, S_k)$ denote the sum $\sum_{T \subseteq
\Pi_{\psi}(\vec{U}), \pi_i(T) = S_i} (-1)^T$. Thus, 
$$
Pr[Q] = \sum_{S_1, \cdots, S_k} N(S_1, \cdots, S_k) Pr[ \bigwedge_{i \in [k]} E(i, S_i)]
$$
The main result of this section is an expression for the quantity
$N(S_1, \cdots, S_k)$. Given a vector $\vec{S} = (S_1, \cdots, S_k)$,
define the signature of $\vec{S}$, denoted $sig(\vec{S})$, as the set
$\{i \mid D_i \neq \emptyset\}$.  

\eat{
Define $\vec{D}^* = (D_1
\cup \{*\}) \times \cdots \times (D_k \cup  \{*\})$. Given any $x \in
\vec{D}^*$,  Let $\vec{D}^*(\psi)$ be the
set $\{x \mid x \in \vec{D}^*, sig(x) \in \psi\}$. Define $$N(\vec{D},
\psi) = \sum_{\{T \subseteq \vec{D}^*(\psi) | \Pi_i(T) = D_i~\text{for
all i}\}} (-1)^{|T|}$$
}

\eat{
\begin{lemma} For any $\vec{D}$ and $\psi$, there exists a function
$f(\psi, sig(\vec{D}))$ that only depends on $\psi$ and $sig(\vec{D})$
such that
$$N(\vec{D}, \psi) = 
    f(\psi, sig(\vec{D}))(-1)^{|D_1| + \cdots + |D_k|} 
$$
\label{lemma:main}
\end{lemma}
}

In an ordered set $(X, <)$, an {\em ideal} is a set of the form $\{x
\mid x \leq a\}$, for a fixed element $a \in X$, which we denote by
$[a]$.  

\begin{lemma} If $[A]$ is an ideal in $\cal{P}(U)$, then $\sum_{\{T \mid T
\in [A]\}} (-1)^T$ = 0 if $A$ is nonempty, and is $1$ if A is empty.
Note that $T \in [A]$ means $T \subseteq A$.
\end{lemma}

For $\vec{S} = (S_1, \cdots, S_k)$, denote 
$$\ND(S) = \sum_{T \subseteq \Pi_\psi(\vec{S})} (-1)^{|T|}$$

\begin{lemma} $N(\vec{S}) = \sum_{\vec{R} \subseteq \vec{S}} (-1)^{|S-R|}
\ND(R)$. Here $\vec{R} = (R_1, \cdots, R_k)$ and $\vec{R} \subseteq
\vec{S}$ means $R_i \subseteq S_i$ for all $i$.
\end{lemma}

\begin{proof} Direct inclusion-exclusion applied to $N(\vec{S})$.
\end{proof}

Define $\UP(\psi) = \{sg \mid sg \subseteq [k], \exists sg' \in \psi s.t. sg' \subseteq
sg\}$.

\begin{lemma} 
\begin{enumerate}
\item If $sig(\vec{R}) \in \UP(\psi)$, then $\ND(\vec{R}) = 0$.
\item If $sig(\vec{R}) \not\in \UP(\psi)$, then $\ND(\vec{R}) = 1$.
\end{enumerate}
\end{lemma}

\begin{proof} Follows from the fact that $sig(\vec{R}) \in \UP(\psi)$
iff $\Pi_\psi(\vec{R}) \not= \emptyset$ and from the fact that
$\ND(\vec{R})$ sums $(-1)^{|T|}$, where $T$ ranges over the ideal
defined by $\Pi_\psi(\vec{R})$.
\end{proof}

Hence, $N(\vec{S}) = (-1)^{|\vec{S}|} \sum_{\{\vec{R} \subseteq
\vec{S}: sig(\vec{R}) \not\in \UP(\psi)\}} (-1)^{\vec{R}}$.

Let $sg$ be a signature, i.e. $sg \subseteq [k]$. Denote the quantity
$M(S,sg) = \sum_{R \subseteq S: sig(R) =sg} (-1)^R$.  Thus, we have:

$$N(S) = (-1)^S * \sum_{sg \not\in \UP(\psi)} M(S,sg)$$

For a signature $sg' \subseteq [k]$, denote:
$$\MD(S,sg') = \sum_{R \subseteq S: sig(R) \subseteq sg'} (-1)^R$$

\begin{lemma} $M(S,sg) = \sum_{sg' \subseteq sg} (-1)^|sg-sg'| *
\MD(S,sg')$
\end{lemma}

\begin{proof} Again inclusion/exclusion formula applied to the set sg.
\end{proof}

\begin{lemma} 
\begin{enumerate}
\item If $sg' \cap sig(S) \neq \emptyset$, then $\MD(S,sg') = 0$.
\item If $sg' \cap sig(S) = \emptyset$, then $\MD(S,sg') = 1$.
\end{enumerate}
\end{lemma} 

\begin{proof} Follows from the fact that the set $\{R \mid R \subseteq S, sig(R)
\subseteq sg'\}$ is an ideal, and it is nonempty iff $sg' \subseteq sig(S)$.
\end{proof}

Next, we manipulate the expression $M(S, sg)$ as follows.
We have $M(S,sg) = (-1)^{sg} M'(S,sg)$,  where:
\begin{eqnarray*}
    M'(S,sg) &=& \sum_{sg' \subseteq sg} (-1)^{sg'} \MD(S,sg')\\ 
             &=& \sum_{sg' \subseteq sg, sg' \cap sig(S) = \emptyset} (-1)^{sg'} \\
             &=& \sum_{sg' \subseteq (sg - sig(S))} (-1)^{sg'}
\end{eqnarray*}

This is a sum over the ideal generated by $sg - sig(S)$. This ideal
contains only the empty set when $sg \subseteq sig(S)$, hence:

\begin{lemma}
\begin{enumerate}
\item If $sg \subseteq sig(S)$, then $M(S,sg) = (-1)^{sg}$
\item If $sg \not\subseteq sig(S)$, then  $M(S,sg) = 0$.
\end{enumerate}
\end{lemma}

Hence,
\begin{eqnarray*}
N(S) &=& (-1)^S * \sum_{sg \not\in \UP(\psi)} M(S,sg)\\
     &=& (-1)^S * \sum_{sg \not\in \UP(\psi), sg \subseteq sig(S)} (-1)^sg\\
     &=& (-1)^S * \sum_{sg \subseteq sig(S)} (-1)^{sg} - (-1)^S * \sum_{sg \in \UP(\psi), sg \subseteq sig(S)} (-1)^{sg}\\
     &=& -(-1)^S * \sum_{sg \in UP(\psi), sg \subseteq sig(S)} (-1)^{sg}
\end{eqnarray*}

The last equality holds because we assume $sig(S) \neq \emptyset$,
hence $sg \subseteq sig(S)$ is a non-empty ideal.

\begin{theorem} $N(S) = -(-1)^S * \sum_{sg \in \UP(\psi), sg \subseteq
sig(S)} (-1)^{sg}$.
\end{theorem}

Next, assume that $\UP(\psi)$ is generated by the set $\psi=\{\phi_1,
\cdots, \phi_p\}$, where each factor $\phi_i$ is a subset of $[k]$.
Then we apply inclusion exclusion to (N6):

$N(S) = -(-1)^S \sum_{G \subseteq [p]} (-1)^{|G-1|} \sum_{\cup_{i \in G}
\phi_i \subseteq sg \subseteq sig(S)} (-1)^{sg}$.

In the inner sum $sg$ ranges over the interval $[\cup_{i \in G} \phi_i,
sig(S)]$, hence the sum is $(-1)^{sig(S)}$ when $\cup_{i \in G} =
sig(S)$ and 0 otherwise.  It follows:

\begin{theorem} 
$N(S) = (-1)^S (-1)^{sig(S)} \sum_{sig(G) = sig(S)} (-1)^G$.
\end{theorem}

\section{Proof of the Dichotomy Theorem} 
\subsection{Unifiers} 

In this section, we define a set ${\cal H}(q)$, called the set of
hierarchical unifiers of $q$, by starting from the factors of $q$ and
unifying them in certain way.

\begin{definition} \em ({\it Hierarchical join predicate}) Let $q_1$
and $q_2$ be two strict hierarchical queries with disjoint sets of
variables and let $g_1 \in \mbox{\it subgoals}(q_1)$ and $g_2 \in
\mbox{\it subgoals}(q_2)$ be any two sub-goals that are unifiable.
Thus, $g_1$ and $g_2$ have same arity, say $a$. Let $m_u : Vars(g_1)
\rightarrow Vars(g_2)$ be the most general unifier of $g_1$ and $g_2$,
which is a bijection.  Let $x_1 \sqsubseteq \cdots \sqsubseteq x_a$ be
all the variables in $g_1$ and $y_1 \sqsubseteq \cdots \sqsubseteq
y_a$ be all the variables in $g_2$.  Let $w$ be the largest integer
such that $m_u(x_i) \equiv y_i$ for $1 \leq i \leq w$. A {\em
hierarchical join predicate} between $q_1$ and $q_2$ is the set $\{
(x_i, m_u(x_i)) \mid 1 \leq i \leq w\}$
\end{definition}

\begin{definition} \em ({\it Hierarchical Unifier}) Let $q_1$ and
$q_2$ be two strict hierarchical queries with disjoint sets of
variables and let $jp$ be some hierarchical join predicate between
them. A hierarchical unifier of $q_1$ and $q_2$ is a query obtained by
considering
$$q_u \leftarrow q_1, q_2, \bigwedge_{(x_i, x_j) \in jp} (x_i = x_j)$$
and removing all $=$ predicates by substituting.
\end{definition}

\begin{lemma} Let $q_u$ be a hierarchical unifier of two strict
hierarchical queries $q_1$ and $q_2$. Then, $q_u$ is a strict
hierarchical query.
\label{lemma:a}
\end{lemma}
\begin{proof} TBD.
\end{proof}

\noindent The above result justifies the name "hierarchical unifier", because
such unifiers are always hierarchical.
Next we define a set ${\cal H}(q)$, called the set of hierarchical
unifiers of $q$, along with a function $\factors$ from ${\cal
H}(q)$ to subsets of ${\cal F}(q)$. They are constructed inductively
as follows:
\begin{enumerate} 
\item For each $q \in {\cal F}(q)$, add $q$ to ${\cal H}(q)$ and let
$\factors(q) = \{q\}$.
\item If $q_1, q_2$ are in ${\cal H}(q)$, and $q_u$ is their hierarchical
unifier, add $q_u$ to ${\cal H}(q)$ if it is not logically equivalent to
any existing query in ${\cal H}(q)$. Also, define
$\factors(q_u)$ to be $\factors(q_1) \cup \factors(q_2)$
\end{enumerate}

\begin{lemma} The set ${\cal H}(q)$ is finite.
\end{lemma}
\begin{proof} All queries in ${\cal H}$ are hierarchical by
Lemma~\ref{lemma:a}. There are only finitely many hierarchical queries
up to equivalence on a given set of relations and given set of
constants. [[Expand this proof]].
\end{proof}

\subsection{The Polynomial Time Algorithm}

Let ${\cal H}^*(q)$ be the subset of ${\cal H}(q)$ containing queries
which are either inversion-free or in ${\cal F}(q)$.

\eat{
The PTIME algorithm for query computation proceeds in 4 steps: (i)
expand the query in terms of the hierarchical unifiers using inclusion
exclusion (ii) add all possible independence predicates to each
remaining signature (iii) apply erases to get rid of bad signatures
(iv) convert the summation back into an (inversion-free) conjunctive
query.
}

\subsubsection{Query expansion}

Let ${\cal H}^*(q) = \{qh_1, qh_2, \cdots, qh_k\}$. Define 
$$\psi = \{S \mid S \subseteq [k],  qc_i \subseteq (\bigcup_{i \in S}
\factors(qh_i))~\text{for some $qc_i \in {\cal C}(q)$}\}$$
Thus, $\psi$ contains all combinations of hierarchical unifiers that
make $q$ true. Let $\factors(\psi)$ be the minimal elements of $\psi$.

\begin{lemma} With $\psi$ as defined above,
$$q \equiv \bigvee_{S \in \psi} \bigwedge_{i \in S} qh_i$$
\end{lemma}

\begin{proof} The $\Leftarrow$ direction is obvious from the
definition of $\phi$. For the $\Rightarrow$ direction, consider any
mapping $\eta$ of $q$ into the database. Consider the factor corresponding to
that mapping and the set of its connected components. This set is in
in $\phi$ and hence $\bigvee_{S \in \psi} \bigwedge_{i \in S} qh_i$ is
true on the database.
\end{proof}

We then apply the generalized inclusion-exclusion formula from
Sec~\ref{sec:pie} to obtain:
$$Pr[q] = \sum_{G \subseteq \factors(\psi)} (-1)^{|G| + |sig(G)|} \sum_{ \{T \mid sig(T) = sig(G)\}} (-1)^T Pr[ qh(T)]$$
where $qh(T) = qh_1(\pi_1(T)), qh_2(\pi_2(T)), \cdots, qh_k(\pi_k(T))$.

Define $\co(sg) = (-1)^{|sg|}\sum_{G \subseteq \factors(\psi), sig(G) = sg} (-1)^{|G|}$.
The sum can alternatively be rewritten as:
$$Pr[q] = \sum_{sg \subseteq [K]} F(sg)$$
where 
$$F(sg) =  \co(sg) \sum_{\{T \mid sig(T) = sg\}} (-1)^{|T|}
Pr[qh(T)]$$

\subsubsection{Adding Independence Predicates}

Let $\bar{x_1}, \cdots, \bar{x_k}$ be the set of variables of $qh_1,
\cdots, qh_k$. Define new relational symbols $S_1, \cdots, S_k$ where
the arity of $S_i$ equals $|\bar{x_i}|$. Given any join predicate $jp$
between $qh_i$ and $qh_j$, consider the following conjunctive query:
$$q_{jp}() \rightarrow S_i(\bar{x_i}), S_j(\bar{x_j}),
\bigwedge_{(x,y) \in jp} (S_i.x = S_j.y)$$

Given any set $T = (T_1, \cdots, T_k)$, let $q_{jp}(T)$ be the predicate
which is true if $q_{jp}$ is true when evaluated on $T$, i.e. by
setting $T_i$ to be the instance of $S_i$.

An {\em independence predicate} is simply the negation of a join
predicate. Let $Q_\ip$ be the set of all independence predicates,
i.e., $Q_\ip = \{ {\texttt not}(q_{jp}) \mid q_{jp} \in Q_\jp\}$. 

We divide the join predicates into two disjoint sets, {\em trivial}
and {\em non-trivial}. A join predicate between factors $h_i$ and
$h_j$ is called trivial if the join query is equivalent to either
$h_i$ or $h_j$, and is called non-trivial otherwise. We write
$\IP(\C^*)$ as $\IP^n(\C^*) \wedge \IP^t(\C^*)$, where $\IP^n(\C^*)$
is the conjunction of $\texttt{not}(jp)$ over all non-trivial join
predicates $jp$, and $\IP^t(\C^*)$ is the conjunction over all trivial
join predicates.

For a signature $sg$, let $\IP^n(sg)$ denote the subset of $Q_\ip$
consisting of independence predicates between all $S_i$ and $S_j$ such
that $i,j \in sg$. Let $\pi$ be a function that maps each signature
$sg$ to a set of independence predicates $\pi(sg) \subseteq \IP^n(sg)$.
Denote:

Let $\pi$ be any predicate on $T$, i.e. a query over the relations
$S_1, \cdots, S_k$. Define
$$sum(\pi) = \sum_{T : \pi(T)} N(sig(T)) (-1)^T Pr[qh(T)]$$

Thus, the probability of $q$ is simply $sum(\emptyset)$, where
$\emptyset$ is the predicate that is identically true. Define $\IP^n$ to
be the conjunction of all independence predicates between queries in
${\cal H}^*(q)$, i.e. $\IP^n = \bigvee_{jp} {\texttt not}(jp)$, where $jp$ ranges
over all join predicates between all $q_i,q_j \in {\cal H}^*(q)$.

We will prove that when the query satisfies the \PTIME\ conditions, then
$sum(\emptyset) = sum(\IP^n)$.

\begin{definition} \em ({\em Eraser}) Let $qh_i$ and $qh_j$ be any two
strict hierarchical queries in ${\cal H}^*(q)$ and let $q_{ij}$ be
their unifier corresponding to some join predicate $jp$.  An {\em
eraser} for the unifier $q_{ij}$ is a set of queries $E \subseteq
{\cal H}^*(q)$ such that:
   \begin{enumerate}
       \item For all $q \in E$, $q \rightarrow q_{ij}$
       \item For all $sg \subseteq [k]$, $N(sg \cup \{i,j\}) = N(sg
       \cup \{i,j\} \cup \{k \mid qh_k \in E\})$.
    \end{enumerate}
\end{definition}

\begin{theorem} Suppose for every $q_i,q_j,q_{ij}$ such that $q_i,q_j
\in {\cal H}^*(q)$ and $q_{ij}$ is a hierarchical unifier of $q_i$ and
$q_j$, either $q_{ij} \in {\cal H}^*(q)$ or it has an eraser. Then,
$sum(\emptyset) = sum(\IP^n)$.
\label{th:add-ip}
\end{theorem}

We will prove Theorem~\ref{th:add-ip} in the rest of this section.

\eat{
Define a set of queries $Q_S$ on $S_1, \cdots, S_k$ inductively as
follows:
\begin{enumerate}
\item $S_i \in Q_S$ for $1 \leq i \leq k$
\item For each $q \in Q_S$ and a join predicate $q_{jp}$ such that
both contain a subgoal $S_i$, let $q_2$ be the query obtained by
unifying the two subgoals. Then, add $q_2$ to $Q_S$. We say that $q_2$
{\em extends} $q$.
\end{enumerate}
}

Let $N$ be the size of the domain for the database. Let ${\cal S}$
denote the vocabulary $S_1, \cdots, S_k$. Let ${\cal Q}_{N,{\cal
S}}(k)$ be the set of conjunctive queries of arity $k$ over ${\cal S}$
that are equivalent on domain of size $N$. For each $q \in {\cal
Q}_{N,{\cal S}}(k)$, define the following 
$$q^* = \exists \bar{x}. q(\bar{x}) \wedge(\bigwedge_{\{q' \mid q'\in
{\cal Q}_{N,{\cal S}}(k), q'~\mbox{\small contains}~ q\}}
{\texttt not}(q'(\bar{x}))$$
Let ${\cal Q}^*_{N,{\cal S}}(k) = \{q^* \mid q \in {\cal Q}_{N,{\cal
S}}(k)\}$ and let ${\cal Q}^*_{N,{\cal S}} = \cup_{k \geq 0} {\cal
Q}^*_{N,{\cal S}}(k)$. Each of the query in ${\cal Q}^*_{N, \cal S}$
is Boolean, hence it contains only finitely many queries up to
equivalence on domains of size $N$, which we denote $\{qs^*_1, qs^*_2, \cdots,
qs^*_t\}$. For each $qs^*_i$, $qs_i$ denotes the conjunctive query
which is the positive part of $qs^*_i$.

\eat{
Given any query $q$ over ${\cal S}$ and any join predicate query
$q_{jp}$ such that both contain a relation $S_i$, let $q'$ be the
query obtained by unifying the two $S_i$ sub-goals. We say that $q'$
extends $q$.

For each $qs_i$, consider the following query 

$$qc_i = qs_i \wedge (\bigwedge_{\{q \mid q ~\mbox{\small contains}~ qs_i\}} {\texttt not}(q))$$
}

We call each such query a {\em cell}. A {\em cell signature} is any
subset of ${\cal Q}^*_{N, \cal S}$.  Given a cell signature $csig$, it
defines the following query 
$$(\bigwedge_{q \in csig} q) \wedge (\bigwedge_{q \not\in
csig} {\texttt not}(q))$$

Given a set $T$, we say $T \models csig$ if $T$ satisfies the query
defined by $csig$. The cell signatures partition the sets of all $T$.
Thus, we have

\begin{eqnarray*}
Pr[q] &=& \sum_{T} N(sig(T)) (-1)^T Pr[qh(T)]\\
      &=& \sum_{csig} \sum_{\{T \mid T \models csig\}} N(sig(T)) (-1)^T Pr[qh(T)]
\end{eqnarray*}

We say that a cell signature {\em csig} contains a join predicate if
their is a cell $qs^*_i \in csig$ and a join predicate query $q_{jp}$
such that $qs^*_i \subseteq q_{jp}$.

\begin{lemma} Let $q$ be the union of all cell signatures that do not
contain any join predicate. Then $T \models q$ iff $T$ satisfies all
the independence predicates.
\end{lemma}

\begin{proof}
\end{proof}

To prove Theorem~\ref{th:add-ip}, we only need to show that the total
contribution of all cell signatures that contain at least one join
predicate is 0. We will show this by grouping cell signatures into
groups of three.

Let $F(csig)$ denote the quantity $\sum_{T \mid T \models csig}
N(sig(T)) (-1)^T Pr[qh(T)]$. Let $qh_i, qh_j$ be any two hierarchical
queries with unifier $q_u$ corresponding to the join predicate $jp$.
$q_{jp}$ is the join predicate query on the {\cal S} vocabulary. 
Let $E = \{qh_{l_1}, \cdots, qh_{l_m}\}$ be its eraser. Thus, there is
a mapping $h : qh_{l_1}, \cdots, qh_{l_m} \rightarrow q_u$. Let
$q_{jp,E} = h(S_{l_1}), \cdots, h(S_{l_m}), q_{jp}$.

Let $qs_m$ be any query that contains $q_{jp}$ but not $q_{jp,E}$.
Thus, there is a mapping $g : q_{jp}$ to $qs_m$. Let $f = h \circ g$
and let $qs_m' = f(S_{l_1}), \cdots, f(S_{l_m}), qs_m$.

Let $csig_0$ be any subset of $Q_S$ that does not contain $qs_m$ and
$qs_m'$.

\begin{lemma} With $csig_0$, $qs_m$ and $qs_m'$ as defined above, 
$$F(csig_0 \cup \{qs_m\}) + F(csig_0 \cup \{qs_m'\}) + F(csig_0 +
\{qs_m, qs_m'\}) = 0 $$
\label{lemma:triplets}
\end{lemma}

\begin{proof} Consider any $T0$ that satisfies either of the three
cells. Then, $T0$ satisfies the query $qs_m$ (note that $qs_m'$
contains the query $qs_m$). Let $H_{l_i}$ be the set of tuples
obtained for $S_{l_i}$ from $qs_m(T0)$ using the mapping $f$.  

Let $T'$ be obtained from $T0$ by removing tuples $H_{l_i}$ from
$T_{l_i}$ for all $i$ in the eraser. Now we fix $T'$ and look at all
the $T$ satisfying either of the three cells and which gives rise to
the same $T'$. Every such $T$ is obtained by adding some subset of
$H_{l_i}$ to $T'_{l_i}$. 

Claim: Every possible $T$ obtained from $T'$ by adding some subset of
$H_{l_i}$ to $T'_{l_i}$ satisfies one of the three cells.

Further, for a fixed $T'$, each $T$ gives rise to the same query
$qh(T)$. Thus, when we sum over all such $T$, we get an ideal which is
0. Summing over all $T'$, we get that the total contribution of the
three cells is 0
\end{proof}

\begin{lemma} The set of all cell signatures that contain at least one
join predicate can be partitioned into groups of three of the form in
Lemma~\ref{lemma:triplets}.
\end{lemma}

\begin{proof} Each triplet is defined by (i) a join predicate
$q_{jp}$ with an eraser $E$, (ii) a pair of queries $qs_a$ and $qs_b$
where $qs_a$ contains $q_{jp}$ but not $q_{jp,E}$ and $qs_b$ is
obtained from $qs_a$ by attaching $E$, and (iii) a subset of cells
$csig_0$. The triplet is then given by: $csig_0 \cup \{qs_a\}$,
$csig_0 \cup \{qs_b\}$ and $csig_0 \cup \{qs_a, qs_b\}$.

Now, given any $csig$ containing a join predicate, define $q_{jp}$,
$qs_a$, $qs_b$ and $csig_0$ as follows.  Order the set of all join
predicates and the set of cells and pick a canonical eraser for each
join predicate. Let $q_{jp}$ be the smallest join predicate in $csig$.
Let $E$ be the canonical eraser for $qs_i$ and let $q_{jp,E}$ be the
query as described above.

Let $qs_m$ be the smallest cell in $csig$ that contains $q_{jp}$. If
$qs_m$ does not contain $q_{jp,E}$, let $qs_a = qs_m$ and define
$qs_b$ appropriately. Note that $csig$ may not contain $qs_b$. If
$qs_m$ contains $q_{jp,E}$ let $qs_b = qs_m$ and define $qs_a$
appropriately. Again, $csig$ may not contain $qs_a$. Let $csig_0$ be
all the cells in $csig$ except $qs_a$ and $qs_b$.  This defines the
triplet for $qs_m$. 

Claim: every cell signature containing a join predicate belongs to a
unique triplet.

This follows from Lemma~\ref{lemma:nocycles}
\end{proof}

\begin{lemma} Let $q_i$ and $q_j$ be two queries with a join predicate
$q_u$ that has an inversion. Suppose $E$ is an eraser for $q_u$, such
that there is a mapping $h : E \rightarrow q_u$. Then, for any $q_l
\in E$, $h(q_l),q_i$ is hierarchical.
\label{lemma:nocycles}
\end{lemma}

\begin{proof} Suppose on the contrary there is an inversion between
$R(x), S(x,y) \in q_l$ and $S(x',y'),T(y')$ in $q_i$ such that
$h(x)=x'$, $h(y)=y'$, where $R(x)$ is some subgoal that contains $x$
but not $y$, $S(x,y)$ is some subgoal containing both $x$ and $y$,
$S(x',y')$ is a subgoal containing both $x'$ and $y'$ and $T(y')$
is a subgoal containing $y'$ but not $x'$.

There are two cases: the join predicate between $q_i$ and $q_j$
does not touch variable $x'$ in $q_j$. Then, no subgoal of $q_i$ in
$q_u$ contains the variable $x'$. So, $h$ maps $R(x)$ to some
subgoal in $q_j$ itself. Thus, $q_j$ is not hierarchical, which is a
contradiction.

Hence, the join predicate between $q_i$ and $q_j$ uses the variable
$x'$. It also uses $y'$ because $x' \sqsubseteq x$. Now, since $q_l$
and $q_i$ have inversion, there must be an eraser $E'$ that has a
mapping to $h(q_l),q_i$. This eraser only uses a portion of the
partial unifier of $q_i,q_j$, hence there is a mapping $E' \rightarrow
q_u$.
\end{proof}
\subsubsection{Change of Basis}

We have
$$Pr[q] = \sum_{T : \IP^n(T)} N(sig(T)) (-1)^T Pr[qh(T)]$$ 
For each $i$,
we expand $qh_i(T_i)$ into the relations it contains. We group all the
$T$ that result in the same $qh(T)$. 

Each $qh_i$ is a connected hierarchical query. Let $\sqsubset$ be the
hierarchy relation on $Vars(qh_i)$. Define a {\em hierarchy tree} for
$qh_i$ as follows. The nodes of the trees are certain subsets of
$Vars(qh_i)$. For each subset of the set of subgoals of $qh_i$, there
is a node in the hierarchy tree consisting of the intersection of
variables of those subgoals. A node $n$ is a child of $n'$ if $n
\subset n'$ and there is no $n''$ such that $n \subset n'' \subset n'$.

For each node in the hierarchy tree of $qh_i$, we define a new
relational symbol whose attributes are the variables in that node. Let
${\cal S}^i = \{S^i_0, S^i_1, \cdots\}$ be the set of new relational
symbols and let $\set{X^i_0, X^i_0, \cdots}$ be the corresponding sets of
variables.

Consider any vector $U_i = (U_{i_1}, U_{i_2}, \cdots)$, where $U_{i_j}
\subseteq A^{Arity(S^i_j)}$. We say that $T \models U$ if for all
$i,j$, $U_{i_j}$ is the projection of $T_i$ on the variables $X^i_j$.
Define $F_i(U_i) = \prod_{j} \prod_{g \mid Vars(g) = X^i_j}
Pr[qh_i(U_{i_j})]$ and let $F(U) = F_1(U_1) \times \cdots \times
F_k(U_k)$. Then, if $T \models U$ and $T$ satisfies all the
independence predicates, we have $Pr[qh(T) = F(U)$.

We rewrite $Pr[q]$ as
$$Pr[q] = \sum_{U} \sum_{\{T \mid U \models T, \IP^n(T)\}} N(sig(T)) (-1)^T Pr[qh(T)]$$ 
The signature of $T$ can be determined by the signature of $U$ in
straightforward way, and we write $N(sig(T))$ as $N(sig'(U))$. Also,
we write $Pr[qh(T)]$ as $F(U)$. We have
$$Pr[q] = \sum_{U} N(sig'(U)) F(U) \sum_{\{T \mid U \models T, \IP^n(T)\}} (-1)^T$$ 
Next, we note that $\IP^n(T)$ is independent of $T$ for a given $U$, and we move
the independence predicates to $U$ as follows: For each independence
predicate between sub-goal $g_1$ of $qh_{i_1}$ and sub-goal $g_2$ of
$qh_{i_2}$, we add the independence predicate ${\texttt not}
(S^{i_1}_{j_1}(\bar{x}), S^{i_2}_{j_1}(\bar{x}))$, where
$S^{i_1}_{j_1}$ is the relation corresponding to $Vars(g_1)$ and
$S^{i_2}_{j_2}$ is the relation corresponding to $Vars(g_12$. Let
$\IP^n(U)$ be the conjunction of all such predicates. Then,
$$Pr[q] = \sum_{\{U \mid \IP^n(U)\}} N(sig'(U)) F(U) \sum_{\{T \mid U \models T\}} (-1)^T$$ 

Not all possible $U$ have a possible $T$. For instance, if relations
$S^i_{j_1}$ and $S^i_{j_2}$ share a set of variable $X$, then $U$ must
have $\pi_{X}(S^i_{j_1}) = \pi_{X}(S^i_{j_2})$. We use the hierarchy
tree to determine when a $U$ has a possible $T$. For each $S^i_{j_1}$
and $S^i_{j_2}$ such that $S^i_{j_1}$ is a child of $S^i_{j_2}$,
define the predicate $S^i_{j_1} = \pi_X(S^i_{j_2}$, where $X$ is the
set of variables in  $S^i_{j_1}$.  Let $\phi$ be the conjunction of
all such predicates on $U$.

\begin{lemma} Let $f(U^i_j)$ be a function which is $(-1)^{|U^i_j|}$
if $S^i_j$ has even number of children in the hierarchy tree of $qh_i$
and 1 otherwise. Then,
$\sum_{\{T \mid U \models T\}} (-1)^{|T|} = \prod_{i,j}
f(U^i_j)$ if $U$ satisfies $\phi$ and 0 otherwise.
\end{lemma}

Using the above lemma, we get
\begin{eqnarray*}
Pr[q] &=& \sum_{\{U \mid \IP^n(U), \phi(U)\}} N(sig'(U)) f(U)F(U)\\
      &=& \sum_{sig} N'(sig) \sum_{\{U \mid \IP^n(U), \phi(U), sig(U) = sig\}} f(U)F(U)
\end{eqnarray*}

Next, we add the remaining independence predicates, namely $\IP^t$.
Consider all pairs $S_{j_1}^{i_1}$ and $S_{j_2}^{i_2}$ in the query
that refer to the same predicate and which have not been separated
using $\IP^n$. Fix an ordering on the subgoals of the query, and let
$g(U^i_j)$ be a function which is $(-1)^{|U^i_j|}$ if there are odd
number of subgoal less than $S_j^i$ that need to be separated from
$S_j^i$ and 1 otherwise. Then,
\begin{eqnarray*}
Pr[q] &=& \sum_{sig} N'(sig) \sum_{\{U \mid \IP^n(U), \IP^t(U),
\phi(U), sig(U) = sig\}} g(U)f(U)F(U)
\end{eqnarray*}
We observe that computing the inner sum is equivalent to evaluating
the query $(\IP^n(U) \wedge \IP^t(U) \wedge \phi(U) \wedge sig(U) = sig)$ on a
probabilistic database with schema $S^i_j$ and instance $U^i_j$ and
probabilities given by $Pr[t \in S^i_j] = g(t)f(t)F(t)/(1 + g(t)f(t)F(t))$.

Finally, to evaluate $(\IP^n(U) \wedge \IP^t(U) \wedge \phi(U) \wedge
sig(U) = sig)$, we negate $\IP^n$ and use inclusion-exclusion to
represent it as probabilities of finite number of conjunctive queries
(with negated subgoals due to $\phi$. Each such conjunctive query is
inversion-free [[need to give more details here]], because the $\IP^n
\wedge \IP^t$ part consists of a bunch of join predicates
corresponding to hierarchical unifiers, and the $\phi$ part also
contains the same join predicates (but with negated sub-goals). So the
resulting query is inversion-free and can be evaluated in \PTIME.

\subsection{Hardness Proof}

The main result of this section is that if there is a hierarchical
unifier that contains an inversion but does not have an eraser, then
the query is $\#P$-hard. This shows that the \PTIME\ condition and the
hardness condition complement each other.

\begin{theorem} Let $q_i, q_j \in {\cal H}^*(q)$ and let $q_k$ be
their hierarchical unifier $q_k$ such that
\begin{enumerate}
\item $q_k$ contains an inversion.
\item $q_k$ does not have any eraser.
\end{enumerate}
Then, $q$ is $\#P$-complete.
\end{theorem}

We prove this in the rest of this section. First, we need some
definitions and results.

\begin{definition} \em ({\em Redundent Set of Covers}) A set of
covers $qc_1, \cdots, qc_k$ is {\em strictly redundant} if there
exists a mapping $h: qc \rightarrow qc_1, \cdots, qc_k$, where $qc$ is
not among $qc_1, \cdots, qc_k$. A set of covers is {\em redundant} if
it contains a strictly redundant subset of covers.
\end{definition}

\begin{definition} \em Let $qc_0, \cdots, qc_k$ be a non-redundant set of
covers. Let $qcs \leftarrow qc_0, \cdots, qc_k$ and define the {\em
cover-set} query to be the minimization of $qcs$ : 
$$qcs' = \text{minimize}(qcs) = qc_0', \cdots, qc_k'$$
where each $qc_i'$ is a subset of subgoals of $qc_i$.  Denote the
{\em inclusions} and the {\em projection} homomorphisms:
\begin{eqnarray*}
  in_i &:& qc_i \rightarrow qcs  ~~~i=0,1,\cdots, k\\
  in   &:& qcs' \rightarrow qcs\\
  pr   &:& qcs  \rightarrow qcs'
\end{eqnarray*}
Note that $pr \circ in$ is the identity mapping on $qcs'$.
\end{definition}

\begin{definition} The mappings $h_i : q \rightarrow qcs'$ obtained by
composing $h :  q \rightarrow qc_i$ (the cover mapping), $in_i$ (the
$i^{th}$ inclusion) and $pr$ (the projection) are called {\em
canonical mappings}.
\end{definition}

\begin{lemma} If $F$ is a non-redundant set of covers, then every
mapping from $q$ to the cover-set query of $F$ is canonical upto
isomorphism.
\end{lemma}

\begin{definition} \em ({\em Extension}) Let $qh \in {\cal H}(q)$ be
any hierarchical unifier with $\factors(qh) = \{qf_1, \cdots, qf_k\}$.
Let $qc_1, \cdots, qc_k$ be a multiset of covers such that $qc_i$ contains
the factor $qf_i$. An {\em extension} of $qh$ is a query $qce'$
obtained by minimizing $qce = qc_1, \cdots, qc_k, qh$. Define the
inclusion homomorphism $in : qce' \rightarrow qce$, the $i^{th}$
inclusion homomorphism $in_i : qc_i \rightarrow qce$ and the
projection homomorphism $pr : qce \rightarrow qce'$ in the natural
way. Also, define canonical mappings for extensions as we defined it
for cover-sets above.
\end{definition}

Now some hardness results.

\begin{lemma} Let $C = \{qc_1, \cdots, qc_k\}$ be a non-redundant set
of covers such that their cover-set $qcs$ has an inversion. Then, $q$
is $\#P$-hard.
\end{lemma}

\begin{proof} Without loss of generality, we can assume that for any
proper subset of $C$, the cover-set does not have an inversion
(otherwise we replace $C$ with the smaller set and repeat the
argument). 

Let the inversion in $qcs$ consist of 
$$g_0(x), h_0(x,y), g_1(u_1, v_1), h_1(u_1,v_1), \cdots,
g_{n-1}(u_{n-1}, v_{n-1}), h_{n-1}(u_{n-1}, v_{n-1}), g_{n}(x',y'), h_{n}(y')$$
where subgoals $h_i$ and $g_{i+1}$ refer to the same relation. For
each $qc_i \in C$, define the type of $qc_i$ as the subset of $[n]$
consisting of all $t$ such that the image of $qc_i$ under the $pr$
homomorphism contains the subgoals $g_t,h_t$.

Claim: for each $qc_i$, its type contains at least one $t$ which is
not present in any other type.

This follows from the minimality of the set $F$, because if $qc_i$
does not contribute any unique $t$, then we can remove if from $F$ and
still get an inversion in the cover-set query.

[[Next use the inclusion-exclusion on the types, and argue that
exactly one conjunct of types is \#P-hard (namely, one that contains
all the types. Use this to give a reduction from RSSS..ST query]]

\end{proof}

\begin{lemma} Let $qh \in {\cal H}(q)$ be a hierarchical unifier that
has an extension $qce$ such that all the mappings from $q \rightarrow
qce$ are canonical. Then $q$ is $\#P$-hard.
\end{lemma}

\begin{proof} We use the extension $qce$ to find a non-redundunct set
of covers whose cover-set has an inversion.

Let $\{qc_1, \cdots, qc_k\}$ be the set of covers used in the
extension of $qce$. Let $h$ be a mapping that maps each variable that
does not participate in the inversion to a unique contant. Construct a
new set of covers $F' = \{qc_1', \cdots, qc_k'\}$ where $qc_i' =
h(qc_i)$. Note that the resulting queries are indeed covers. We will
show that $F'$ is a non-redundunct set of covers whose cover-set has
an inversion.

It is easy to see that the cover-set of $F'$ is precisely the query
$h(qce)$. Since $h$ does not touch the variables that participate in
the inversion, $h(qce)$ also contains an inversion, and so does
$h(qce)$. To prove that $F'$ is non-reduntant, we note that a
non-canonical mapping from some cover $qc$ to the cover-set of $F'$
gives a non-canonical mapping from a different cover $qc'$ (obtained
by replacing the new constants back by varaibles) to the extension
$qce$. This is a contradiction, hence all mappings into the cover-set
of $F'$ are canonical. So $F'$ is non-reduntant.
\end{proof}

\begin{lemma} Let $qh \in {\cal H}(q)$ be a hierarchical unifier that
does not have an eraser. Then, there is an extension $qce$ such that
all the mappings from $q \rightarrow qce$ are canonical.
\end{lemma}

\end{document}